\documentclass[journal]{new-aiaa}
\usepackage[utf8]{inputenc}

\usepackage{graphicx}
\graphicspath{{figures/}}
\usepackage{amsmath}
\usepackage{bm}
\newcommand*\dif{\mathop{}\!\mathrm{d}}
\usepackage[version=4]{mhchem}
\usepackage{siunitx}
\usepackage{longtable,tabularx}
\usepackage[usenames,dvipsnames]{xcolor}    
\usepackage[]{hyperref}
\hypersetup{
  colorlinks=true,
  linkcolor=black,
  urlcolor=black,
  citecolor=black,
}

\definecolor{myblue}{HTML}{4C72B0}
\definecolor{myred}{HTML}{C54E52}
\definecolor{mygreen}{HTML}{56A968}

\title{Field Inversion Machine Learning for Time-Resolved Unsteady Flows in Airfoil Dynamic Stall}

\author{Zilong Li\footnote{PhD Student, Department of Aerospace Engineering, AIAA Student Member.}}
\author{Lean Fang\footnote{Postdoctoral Researcher, Department of Aerospace Engineering, AIAA Member.}}
\author{Anupam Sharma\footnote{Professor, Department of Aerospace Engineering,  AIAA Associate Fellow.}}
\author{Ping He\footnote{Assistant Professor, Department of Aerospace Engineering, AIAA Senior Member. Email: phe@iastate.edu}}
\affil{Iowa State University, Ames, IA 50011}
\begin{document}

\maketitle

\begin{abstract}

While many existing machine learning studies have focused on augmenting Reynolds-averaged Navier–Stokes (RANS) turbulence models for steady or time-averaged unsteady flows, this paper takes a first step toward extending such augmentation to time-resolved unsteady flows. An unsteady field inversion and machine learning (FIML) method is developed, in which a temporally evolving correction field ($\beta$) is incorporated into the production term of a RANS turbulence model. The inverse problem is solved by optimizing the spatial–temporal distribution of $\beta$ to minimize the regularized prediction errors. The resulting optimized $\beta$ field is then used to train a multi-layer neural network that learns the time-dependent relationship between local flow features and $\beta$. The approach is demonstrated using the unsteady flow over a NACA0012 airfoil undergoing dynamic stall. Results show that the unsteady FIML model, trained using only the time series of drag data at a given pitch rate, can accurately reproduce the spatial–temporal evolution of reference drag, lift, pitching moment, surface pressure, and velocity fields at both identical and different pitch rates. The unsteady FIML is integrated into the open-source DAFoam framework, enabling a pathway toward developing accurate and generalizable RANS turbulence models for time-resolved unsteady flows.

\end{abstract}

\section{Introduction}
\label{sec_introduction}
Dynamic stall is a complex, highly unsteady aerodynamic phenomenon that occurs in rapidly manoeuvring aircraft~\cite{brandon1991dynamic}, rotor blades~\cite{ham1968dynamic}, or wind turbines~\cite{fujisawa2001observations,larsen2007dynamic}.
It is characterized by large flow separation and strong adverse pressure gradient, and results in significant variations in drag, lift, and pitching moment that far exceed their static stall counterparts.
Based on the nature of the boundary layer separation preceding stall, ~\citet{mccroskey1981dynamic} classified the airfoil stall into four types: leading edge stall, trailing edge stall, thin airfoil stall, and mixed stall.
Leading-edge stall occurs when the laminar separation bubble (LSB) collapses or an abrupt flow reversal occurs at the leading edge.
Trailing edge stall begins with the onset of flow reversal near the trailing edge.
Thin airfoil stall is characterized by the progressive lengthening of the LSB until it extends over the entire airfoil.
Mixed stall may develop when flow separation develops simultaneously near both the leading and trailing edges or separation originates near the mid-chord.
In airfoil design, the trailing-edge stall is generally preferred because it results in a more gradual reduction in lift compared to other stalls.
Consequently, accurately assessing the effects of adverse pressure gradients on the suction surface and reliably predicting the onset of flow separation are crucial in airfoil dynamic stall analysis.

While early investigations into airfoil stall mechanisms relied on experimental studies~\cite{carr1977analysis,broeren1999flowfield,mulleners2012onset,mulleners2013dynamic,heine2013dynamic,muller2015control,muller2016dynamic}, nowadays, research predominantly utilizes numerical simulations, such as computational fluid dynamics (CFD).
CFD is a powerful tool for analyzing three-dimensional flow fields around airfoils, providing valuable insights into flow physics that support aerospace system designs.
Computational investigations of airfoil stall can be performed at various levels of fidelity, including low-fidelity Reynolds-averaged Navier-Stokes (RANS) simulations~\cite{visbal1990dynamic,ekaterinaris1995numerical}, medium-fidelity large-eddy simulations (LES)~\cite{mary2002large,garmann2011numerical,visbal2011numerical, visbal2017numerical,asada2018large,visbal2018analysis,sharma2019numerical,tamaki2020physics,tamaki2023wall}, and high-fidelity direct numerical simulations (DNS)~\cite{jones2008direct,rodriguez2013direct,hosseini2016direct,rosti2016direct}.
However, the extremely high spatial and temporal resolution~\cite{choi2012grid} required by the medium- and high-fidelity simulations makes them computationally expensive for practical engineering design and optimization problems, in which hundreds of CFD simulations may be needed.
In the foreseeable future, engineering design will probably still rely on RANS simulations with turbulence models, such as the one-equation Spalart-Allmaras (SA)~\cite{spalart1992one} model and two-equations $k-\varepsilon$~\cite{launder1974application} and $k-\omega$~\cite{wilcox1988reassessment} model.
However, the RANS turbulence models have imperfect assumptions and simplifications, often resulting in inaccurate predictions of complex flows, particularly for those involving large flow separation and strong adverse pressure gradients, such as dynamic stall.
\citet{celic2006comparison} reported that the SA model exhibits significant deficiencies in predicting boundary-layer separation under adverse pressure gradients, which in turn limits its capability to accurately capture airfoil stall behavior.
Balancing the computational cost and prediction accuracy for airfoil dynamic stall simulations remains a challenging task.

To address the inaccuracies in RANS models, researchers have employed the machine learning (ML) method to correct the defects in the governing equation of a RANS turbulence model. 
In contrast to traditional, human-intuition-based turbulence model development, ML methods utilize data to expedite the development of accurate and broadly applicable turbulence models.
Field inversion machine learning (FIML) is a model-consistent approach originally proposed by Duraisamy and co-workers~\cite{parish2016paradigm,singh2017machine,singh2016using} and becomes popular in recent years.
One of FIML's key advantages is its integration of a CFD solver during the training phase (field inversion), which ensures consistency between training and prediction phases at the discretized level, thereby enhancing both accuracy and generalizability~\cite{duraisamy2021perspectives}.
Additionally, FIML can use various types of training data, including integrated values, surface variables, and sparse or partial field data.
A brief review of existing FIML research is presented as follows.

\citet{singh2017machine} enhanced the Spalart-Allmaras (SA) turbulence model by training it with experimental lift coefficient data to predict flow separation over airfoils.
The augmented model significantly improved lift prediction accuracy across various angles of attack and demonstrated generalizability to untrained flow conditions, geometries, and solvers.
To improve model consistency, \citet{holland2019field} introduced a coupled field inversion and machine learning framework, combining inference and learning processes in a unified approach.
\citet{he2018data} developed a continuous adjoint method for field inversion, showcasing reasonably good performance across diverse 2D and 3D flow scenarios.
\citet{ferrero2020field} used the wall isentropic Mach number data to augment the SA model for accurate flow predictions in gas turbine cascades.
Their trained model demonstrated good prediction accuracy for various unseen Mach numbers and blade geometries.
\citet{wu2023enhancing} proposed a symbolic regression method to improve the interpretability of FIML results.
Their augmented $k-\omega$ SST turbulence model exhibits reasonably good generalizability for predicting separated flow for various 2D and 3D configurations.
Later, the authors extended their FIML work to consider conditional field inversion~\cite{wu2024development}, nonlocal effects~\cite{wu2025data}, and a ``rubber-band" strategy~\cite{wu2025field}.
Instead of using an adjoint-based approach, \citet{strofer2021dafi} developed an open-source framework that employed the ensemble Kalman filtering (EnKF) method for field inversion.
This approach was further explored by \citet{yang2020improving}, who used the EnKF method to study the laminar-to-turbulent flow transition over airfoils.
They augmented a four-equation $k-\omega-\gamma-A_r$ turbulence model, which demonstrated reasonably good performance in predicting the transition location across various angles of attack.
The EnKF method has also been applied to improve turbulence model predictions for the interaction between the shock wave and boundary layer~\cite{tang2023improvement}, laminar-to-turbulent flow transition in hypersonic boundary layer~\cite{zhang2023improvement}, and flow over a hump~\cite{yi2023improvement}.

Despite the above progress, all the FIML studies cited above have been limited to steady-state flow problems. 
This limitation is largely due to the significant challenges associated with FIML, particularly the need for an adjoint solver to compute gradients for a large number of design variables~\cite{duraisamy2021perspectives}. 
Developing an efficient adjoint solver requires access to source codes, a deep understanding of the CFD solver, and considerable time investment.
Although the EnKF method~\cite{strofer2021dafi} has been proposed as an alternative to adjoint‐based field inversion, achieving comparable accuracy would require substantially higher computational cost than the adjoint approach, especially for unsteady flows.
Recently, \citet{fidkowski2022gradient} demonstrated that a steady-state trained FIML model could improve prediction accuracy for periodic unsteady flow.
He used time-averaged data from periodic unsteady flow (ignoring the unsteady flow time history) and conducted steady-state FIML to correct the turbulence modeling error.
He then used the corrected model to predict unsteady periodic flow and incorporated it into aerodynamic shape optimization.
However, our recent work~\cite{fang2024field} showed that there is no guarantee that a steady-state trained FIML model can accurately predict the time history of general, non-periodic unsteady flow.
In that study, we developed an FIML framework to improve the RANS model for predicting time-resolved unsteady flows over a ramp.

This paper takes a further step by enabling unsteady FIML for the accurate prediction of RANS CFD in time-resolved unsteady flows over a pitching up airfoil.
The proposed framework adds a temporally evolving correction field ($\beta$) into the production term of a RANS turbulence model. 
The inverse problem is then solved by optimizing the spatial–temporal distribution of $\beta$ to minimize the regularized prediction errors. 
The resulting optimized $\beta$ field is then used to train a multi-layer neural network that learns the time-dependent relationship between local flow features and $\beta$. 
We will demonstrate the proposed FIML framework using the time-resolved unsteady flow over a NACA0012 airfoil undergoing dynamic stall.
The FIML framework has been integrated into our open-source CFD-based optimization framework, DAFoam~\cite{he2020dafoam}.
Information regarding the download, installation, and utilization of DAFoam can be accessed on its documentation website at \url{https://dafoam.github.io}.
Examples of DAFoam-based steady and unsteady FIML studies can be found in these publications~\cite{fang2024field,bidar2022opensource,wu2024development,wu2023enhancing,chenyu2024field,li2025field}.

The remainder of the paper is structured as follows. In Section~\ref{sec_method}, we provide detailed explanations of the proposed unsteady FIML framework and its components. The results of the unsteady FIML approach are presented and analyzed in Section~\ref{sec_results}, followed by a summary of our findings in Section~\ref{sec_conclusion}.

\section{Method}
\label{sec_method}

In this section, we provide a detailed description of the proposed unsteady FIML framework, unsteady flow simulations, unsteady adjoint computation, the pitching motion of the NACA0012 airfoil, and the CFD setup. 

\subsection{FIML framework for steady-state and time-resolved unsteady flows}

\begin{figure*}[!t] 
  \centering
  \includegraphics[width=0.98\linewidth]{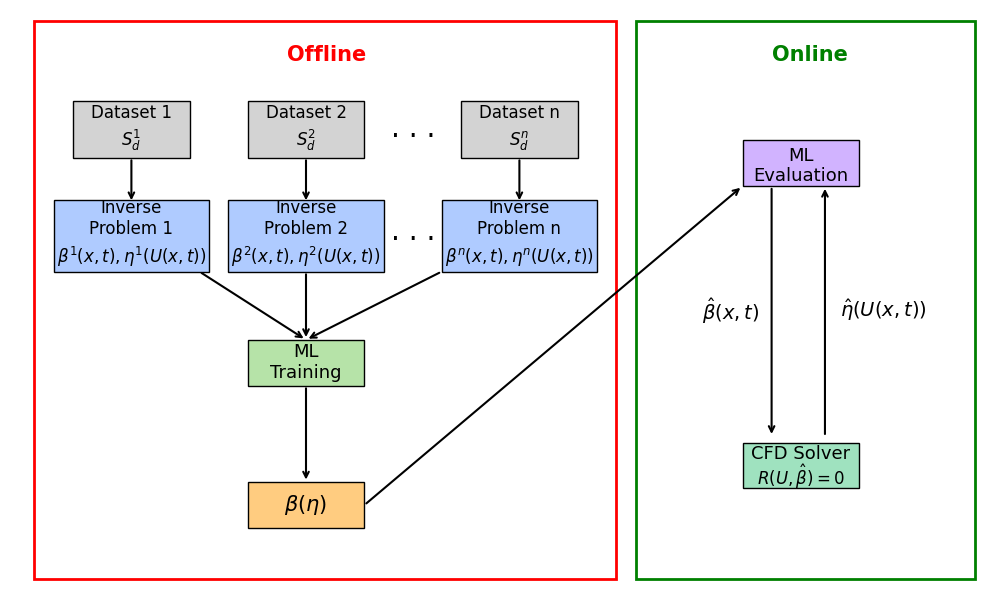}
  \caption{Schematic of field inversion machine learning (FIML) framework for the augmented turbulence modeling.}
  \label{fig_fiml}
\end{figure*}

In this study, we extend the FIML framework to augment time-resolved unsteady flows with moving CFD boundaries. 
A schematic of the FIML framework is shown in Fig.~\ref{fig_fiml}.
As mentioned above, the augmentation field $\beta$ is incorporated in the turbulence production term.
Starting with the reference dataset $S_d$, we will solve an inverse problem, in which we optimize the spatial-temporal distribution of $\beta$ to minimize the prediction error of the RANS model.
In this study, we evaluate both steady and unsteady field inversion strategies; therefore, $\beta$ is a spatial and spatial-temporal field for the steady and unsteady cases, respectively.
For the steady field inversion problem, we minimize the prediction error for the converged steady-state flow fields, whereas for the unsteady field inversion, we minimize the prediction error computed from time-resolved unsteady flow fields.

The augmentation $\beta$ fields are obtained from an ensemble of the inverse problems on different data sets (e.g., $S^1_d$, $S^2_d$, ..., $S^n_d$ in the figure) that are representative of the flow physics to be considered. 
For example, we use data across multiple angles of attack for the steady field inversion and data from selective snapshots of the time-resolved CFD simulations for the unsteady case.
To be useful in predictive simulation, the augmentation field $\beta$ has to build a relationship with flow features $\eta(U(x,t))$, where $U(x,t)$ represents the flow and turbulence variables.
This paper trains a multi-layer neural network (NN) model to construct a function relationship of $\beta(\eta)$.
The FIML process operates as an offline, pre-processing stage, as indicated by the red box in the schematic. 
Once the NN model is trained, the CFD solver continuously provides the flow features $\hat{\eta}(U(x,t)$ to the NN model at each time step or iteration.
The NN model then predicts the corresponding augmentation quantity $\hat{\beta}(x,t)$ and returns it to the CFD solver, where it is incorporated into the turbulence model to enhance the prediction capability.
This corresponding to the online (predictive simulation) stage shown in the green box.

\begin{figure*}[!t] 
  \centering
  \includegraphics[width=0.85\linewidth]{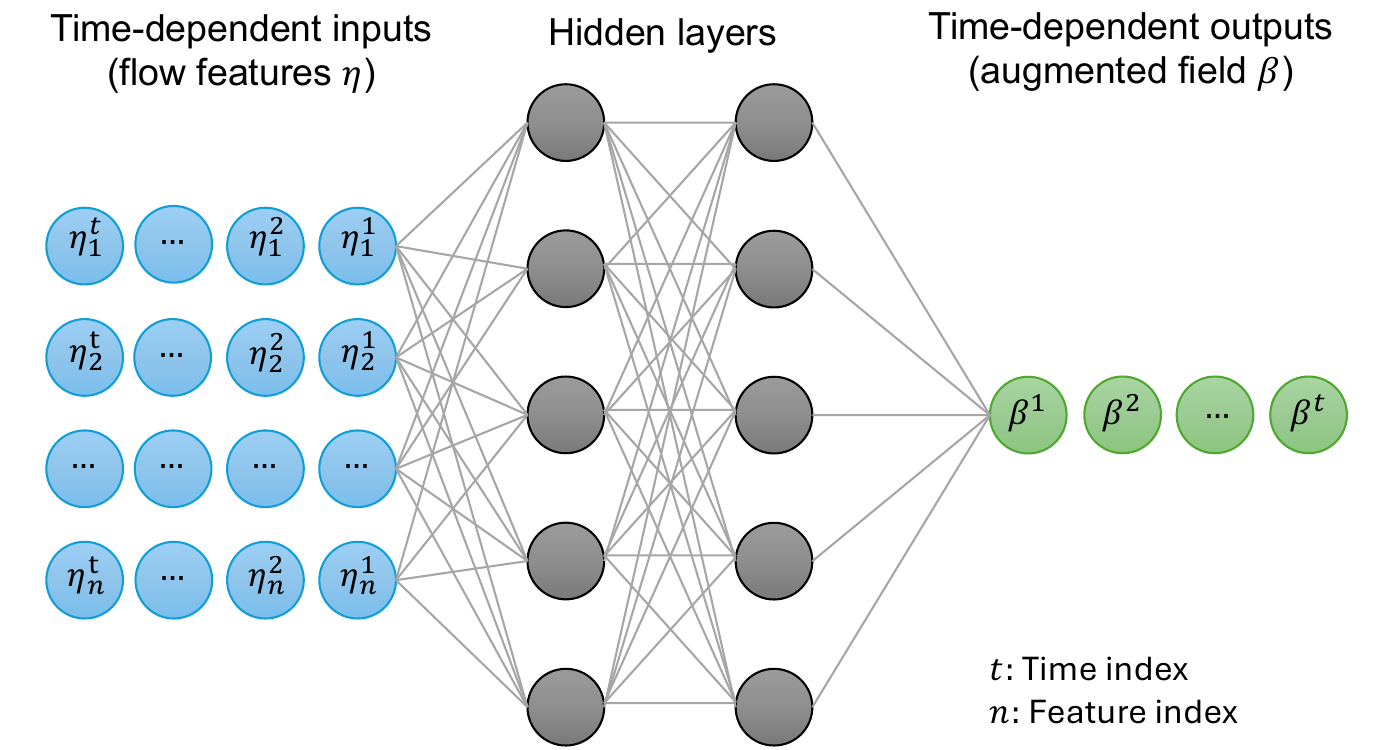}
  \caption{Schematic of the multi-layer neural network (machine learning) model in the offline stage for augmentation field computation.}
  \label{fig_nn_diagram}
\end{figure*}

The neural network (machine learning) model for the augmentation field computation is detailed in Fig.~\ref{fig_nn_diagram}.
The neural network takes local flow features $\eta$ as inputs and outputs the augmentation field $\beta$. 
It consists of an input layer, one or more hidden layers, and an output layer.
For steady-state problems, the inputs and outputs are from one time instance (the final converged state), whereas for unsteady cases, the inputs and outputs are from multiple time instances.
In other words, for unsteady cases, we train only one NN model that can accurately predict $\beta$ for each time step. 

In the NN model, each neuron is computed as a weighted sum of the neurons from the previous layer and passed through a nonlinear activation function

\begin{equation}
\label{eqn_nn}
x^k_i = f_a(b_i + \sum_{j=1}^{N_{k-1}} w_j x_j^{k-1}) ,
\end{equation}
where $w$ and $b$ represent the weights and biases, respectively. 
The superscripts $k$ and $k-1$ denote the current and previous layers, respectively. 
$N_{k-1}$ is the total number of neurons in the $k-1$ layer.
Various options are available for the activation function $f_a$, including tanh, sigmoid, and ReLU. 
In this study, we select the hyperbolic tangent activation function (tanh), which is expressed
as
\begin{equation}
\label{eqn_nn_act}
f_a(y) = \frac{1-e^{-2y}}{1+e^{-2y}} .
\end{equation}

We apply the Keras libraries in TensorFlow to implement the training process of the nerual network model.
We use the Adam optimizer to efficiently find the minimum of a loss function by adaptively adjusting the learning rate for each parameter during model training.
For both the steady- and unsteady-FIML cases, a four-layer neural network architecture is adopted, consisting of an input layer, two hidden layers, and an output layer. 
The network takes four local flow features as inputs and predicts the corresponding augmentation $\beta$ field. 
The primary difference between the two models lies in the size of the hidden layers: the steady-FIML case employs 20 neurons in each hidden layer, whereas the unsteady-FIML case utilizes 100 neurons per layer to accommodate the increased complexity of time-resolved unsteady flow aerodynamics. 
Note that the neuron number is selected through trial and error; increasing the number of neurons does not necessarily yield improved performance and may lead to overfitting or increased computational cost.

\begin{figure*}[!t] 
  \centering
  \includegraphics[width=\linewidth]{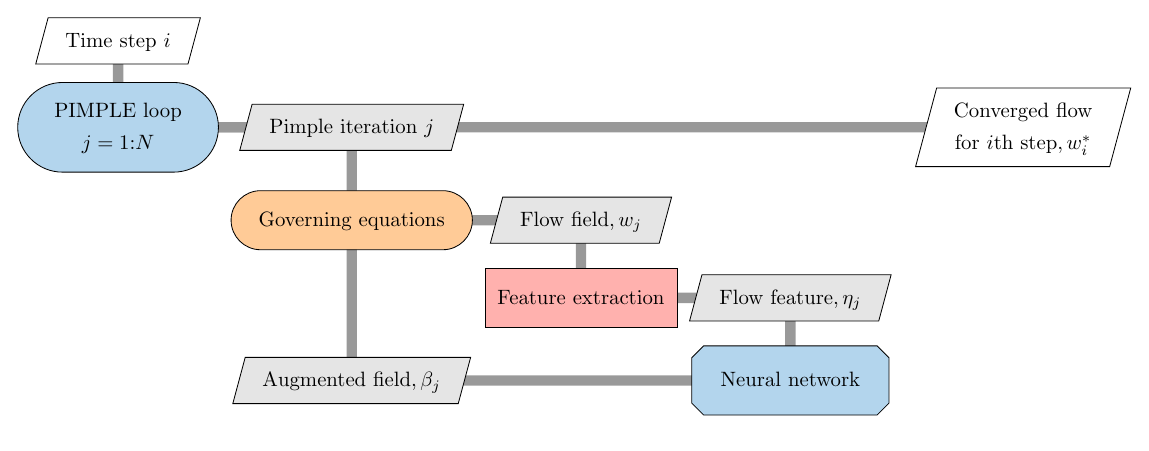}
  \caption{Incorporation of a trained neural network model into the unsteady CFD solver for predictive simulation (online stage).}
  \label{fig_fiml_solver}
\end{figure*}

After the NN model is trained, we proceed to the online predictive simulation, as illustrated in Fig.~\ref{fig_fiml_solver}. 
A feature extraction component calculates local flow features ($\eta$) from the latest flow field ($w$).
The NN model then takes the flow features $\eta$ as inputs to compute the augmentation scalar field $\beta$ for each iteration.
Note that here we reuse the pre-trained neural network model in Fig.~\ref{fig_nn_diagram}.
The evaluated augmentation field $\beta$ will be incorporated into the governing equation of the RANS turbulence model to improve its prediction accuracy.

\subsection{Unsteady flow simulation using the PIMPLE method}
We use DAFoam's DAPimpleDyMFoam to simulate unsteady flow with moving boundaries. DAPimpleDyMFoam is a modified version of OpenFOAM's~\cite{weller1998tensorial} built-in solver \texttt{pimpleFoam} that considers a dynamic mesh. 
It solves three-dimensional, unsteady, turbulent flow governed by the incompressible Navier--Stokes equations:

\begin{equation}
  \label{eq:continuity}
   \nabla \cdot \boldsymbol{U} = 0,
\end{equation}
\begin{equation}
  \label{eq:momentum}
   \frac{\partial \boldsymbol{U}}{\partial t} + ((\boldsymbol{U} - \boldsymbol{U}_{m}) \cdot \nabla) \boldsymbol{U}  +  \nabla p -  \nabla \cdot \nu_\textrm{eff} (\nabla \boldsymbol{U} +[\nabla \boldsymbol{U}]^T) = 0 ,
\end{equation}
where $t$ is the time, $p$ is the kinematic pressure, $\boldsymbol{U}$ is the velocity vector $\boldsymbol{U}=[u, v, w]$,
$\boldsymbol{U}_m$ is the mesh velocity,
$\nu_\textrm{eff}=\nu+\nu_t$ with $\nu$ and $\nu_t$ being the kinematic and turbulent eddy viscosity, respectively.

As previously indicated, we solve the Navier-Stokes equations using the PIMPLE method, which combines features of the PISO and SIMPLE algorithms. 
The procedures involved are briefly outlined below.

First, the momentum equation is discretized, and an intermediate velocity field is solved using either the pressure field obtained from the previous iteration ($p^{t-\Delta t}$) or an initial guess.
For simplicity, we assume the first-order Euler scheme is employed for temporal discretization.

\begin{equation}
  \label{eqn_momentum_predictor}
  a_P \boldsymbol{U}_P^t = - \sum_N a_N \boldsymbol{U}_N^t + \frac{\boldsymbol{U}^{t-\Delta t}_P}{\Delta t} \frac{V^{t-\Delta t}}{V^t} - \nabla p^{t-\Delta t}  = \boldsymbol{H}(\boldsymbol{U}) - \nabla p^{t-\Delta t} ,
\end{equation}
where $a$ is the coefficient resulting from the finite-volume discretization, subscripts $P$ and $N$ denote the control volume cell and all of its neighboring cells, respectively, $\boldsymbol{U}^{t-\Delta t}$ is the velocity from the previous time step, and

\begin{equation}
  \label{eqn_h_term}
  \boldsymbol{H}(\boldsymbol{U}) = - \sum_N a_N \boldsymbol{U}_N^t + \frac{\boldsymbol{U}^{t-\Delta t}_P}{\Delta t} \frac{V^{t-\Delta t}}{V^t}
\end{equation}
represents the influence of velocity from all the neighboring cells and from the previous iteration.
Mesh volumes of the current and previous time steps, $V^t$ and $V^{t-\Delta t}$, are also introduced due to the mesh motion.
A new variable $\phi$ (face flux) is introduced to linearize the convective term, and for a dynamic mesh we use the relative face flux due to the mesh motion:

\begin{equation}
  \label{eqn_conv_term}
  \int_S (\boldsymbol{U} - \boldsymbol{U}_m) \boldsymbol{U}  \cdot \dif \boldsymbol{S}  = \sum_f (\boldsymbol{U}_f - {\boldsymbol{U}_m}_f) \boldsymbol{U}_f \cdot \boldsymbol{S}_f = \sum_f (\phi - \phi_m) \boldsymbol{U}_f
\end{equation}
where the subscript $f$ denotes the cell face, $\phi$ can be obtained from the previous iteration or an initial guess, and $\phi_m$ is the mesh face flux calculated from the mesh motion.
By solving the momentum equation~\eqref{eqn_momentum_predictor}, we obtain an intermediate velocity field that does not yet satisfy the continuity equation.

Next, the continuity equation is coupled with the momentum equation to construct a pressure Poisson equation, from which an updated pressure field is calculated. The discretized form of the continuity equation is given as:

\begin{equation}
  \label{eqn_mass_discretized}
  \int_S \boldsymbol{U} \cdot \dif \boldsymbol{S} = \sum_f \boldsymbol{U}_f \cdot \boldsymbol{S}_f = 0.
\end{equation}
Rather than employing a straightforward linear interpolation, $\boldsymbol{U}_f$ in this equation is determined by interpolating the cell-centered velocity $\boldsymbol{U}_P$, which is derived from the discretized momentum equation~\eqref{eqn_momentum_predictor}, onto the cell face using the following method:

\begin{equation}
  \label{eqn_momentum_interpol}
  \boldsymbol{U}_f = \left (\frac{\boldsymbol{H}(\boldsymbol{U})}{a_P} \right)_f - \left( \frac{1}{a_P} \right)_f (\nabla p)_f .
\end{equation}

This idea of momentum interpolation was initially proposed by Rhie and Chow~\cite{rhie1983numerical} and is effective in mitigating the oscillating pressure (checkerboard) issue resulting from the collocated mesh configuration.
Substituting Eq.~\eqref{eqn_momentum_interpol} into Eq.~\eqref{eqn_mass_discretized}, the pressure Poisson equation can be obtained:
\begin{equation}
  \label{eqn_pressure_possion}
  \nabla \cdot \left( \frac{1}{a_P} \nabla p \right) = \nabla \cdot \left( \frac{\boldsymbol{H}(\boldsymbol{U})}{a_P} \right).
\end{equation}

Solving Eq.~\eqref{eqn_pressure_possion} provides an updated pressure field $p^t$.
The new pressure field $p^t$ is then used to correct the face flux
\begin{equation}
  \label{eqn_phi}
  \phi^t =  \boldsymbol{U}_f \cdot \boldsymbol{S}_f =  \left[ \left (\frac{\boldsymbol{H}(\boldsymbol{U})}{a_P} \right)_f - \left( \frac{1}{a_P} \right)_f (\nabla p^t)_f  \right] \cdot \boldsymbol{S}_f ,
\end{equation}
and velocity field
\begin{equation}
  \label{eqn_momentum_corrector}
   \boldsymbol{U}^t  = \frac{1}{a_P}[ \boldsymbol{H}(\boldsymbol{U}) - \nabla p^t ].
\end{equation}

Here, the $\boldsymbol{H}(\boldsymbol{U})$ term depends on $\boldsymbol{U}$ but has not yet been updated.
To account for this, we need to iteratively solve the Eqs.~\eqref{eqn_h_term} to~\eqref{eqn_momentum_corrector} within the PISO corrector loop. 
In this study, we employ two iterations of the PISO corrector loop.

To close the system and connect turbulent viscosity to the mean flow variables, a turbulence model is required. 
The Spalart-Allmaras (SA) model is used, which solves the following equation:
\begin{equation}
  \label{eq_sa_model}
   \frac{\partial \tilde{\nu}}{\partial t} + \nabla \cdot (\boldsymbol{U} - \boldsymbol{U}_m) \tilde{\nu}  - \dfrac{1}{\sigma} \{ \nabla \cdot [ (\nu+\tilde{\nu}) \nabla \tilde{\nu} ] + C_{b2} |\nabla \tilde{\nu} |^2 \} - \beta C_{b1} \tilde{S} \tilde{\nu} + C_{w1} f_w \left( \dfrac{\tilde{\nu}}{d} \right)^2  =0 .
\end{equation}
where $\tilde{\nu}$ is the modified viscosity, and it is related to the turbulent eddy viscosity as
\begin{equation}
\label{eqn_nutilda}
\nu_t = \tilde{\nu} \frac{\chi^3}{\chi^3+C_{v1}^3}, \quad \chi = \frac{\tilde{\nu}}{\nu} .
\end{equation}
Refer to \citet{spalart1992oneequation} for a more detailed description of the terms and parameters in the SA model.
In our framework, the production term $P$ is multiplied by an augmentation field $\beta$.

In addition to the PISO corrector loop mentioned above, the PIMPLE algorithm iteratively solves Eqs.~\eqref{eqn_momentum_predictor} to~\eqref{eq_sa_model} multiple times until all flow residuals are reduced to a sufficiently low level (PIMPLE corrector loop).
To ensure the stability of PIMPLE, it is necessary to under-relax the solutions of the momentum equation~\eqref{eqn_momentum_predictor}, turbulence equation~\eqref{eq_sa_model}, and the pressure is updated after solving the pressure Poisson equation~\eqref{eqn_pressure_possion}, except for the last PIMPLE corrector loop.
The PIMPLE method allows for the use of relatively large time steps (CFL$>$1). 
While this feature may not significantly accelerate unsteady flow simulations due to the need for multiple PIMPLE iterations per time step, it is highly advantageous for the unsteady adjoint solver. Larger time steps reduce the number of adjoint equations to solve and the amount of intermediate flow data to manage.
In this study, we iterate the PIMPLE corrector loop until flow residuals decrease by 8 orders of magnitude or a maximum of 5 iterations is reached.

\subsection{PIMPLE-Krylov unsteady adjoint formulation for gradient computation}
As previously stated, the PIMPLE flow simulation maintains tight convergence of all flow residuals at each time step.
\begin{equation}
  \label{eqn_xw_to_r_time_series}
  \boldsymbol{R}(\boldsymbol{x},\boldsymbol{w}) = 
  \begin{bmatrix}
 \boldsymbol{R}^1(\boldsymbol{x},\boldsymbol{w}^1,\boldsymbol{w}^0) \\
 \boldsymbol{R}^2(\boldsymbol{x},\boldsymbol{w}^2,\boldsymbol{w}^1, \boldsymbol{w}^0) \\
 \vdots \\
 \boldsymbol{R}^K(\boldsymbol{x},\boldsymbol{w}^K,\boldsymbol{w}^{K-1}, \boldsymbol{w}^{K-2}) \\
\end{bmatrix}
  = \boldsymbol{0} ,
\end{equation}
where the superscript denotes the time step index with $K$ being the total number of time steps, $\boldsymbol{x}\in \mathbb{R}^{n_x}$ represents the design variable vector with $n_x$ being the total number of design variables, $\boldsymbol{w} \in \mathbb{R}^{Kn_w}$ is the state variable vector with $n_w$ being the total number of state variable for each time step, and $\boldsymbol{R}\in \mathbb{R}^{Kn_w}$ is the flow residual vector.
In this study, we employ an implicit second-order time discretization scheme for all time steps except the initial one, where a first-order time scheme is utilized.
In the primal unsteady flow solution, Eq.~(\ref{eqn_xw_to_r_time_series}) is solved in a forward manner to determine the state variable for all time steps, i.e., $\boldsymbol{w}^1, \boldsymbol{w}^2, \ldots, \boldsymbol{w}^K \in \mathbb{R}^{n_w}$.

The objective function \emph{F} depends on both the design variables $\boldsymbol{x}$ and the state variable $\boldsymbol{w}$ solved in Eq.~(\ref{eqn_xw_to_r_time_series}). 
In many applications, including the current study, the objective function \emph{F} can be represented as the average of a time series, i.e.,
\begin{equation}
  \label{eq:obj_func}
  F(\boldsymbol{x},\boldsymbol{w}) = \dfrac{1}{K} \sum_{i=1}^K f^i(\boldsymbol{x},\boldsymbol{w}^i) ,
\end{equation}
where for each $1 \leq i \leq K$, $f^i$ depends only on the design variables $\boldsymbol{x}$ and the corresponding state variable $\boldsymbol{w}^i$ at that specific time step.
This formulation allows the partial derivative $\partial F/\partial \boldsymbol{w}$ to be simplified as:
\begin{equation}
  \label{eq:partial_dfdw_simplify}
 \underbrace{\dfrac{\partial F}{\partial \boldsymbol{w}}}_{1 \times Kn_w} = \ \dfrac{1}{K} \
 [
\underbrace{\dfrac{\partial f^1}{\partial \boldsymbol{w}^1}}_{1 \times n_w},  \
\underbrace{\dfrac{\partial f^2}{\partial \boldsymbol{w}^2}}_{1 \times n_w}, \
\cdots, \
\underbrace{\dfrac{\partial f^K}{\partial \boldsymbol{w}^K}}_{1 \times n_w}
]
.
\end{equation}
For other typical forms of objective functions, such as the variance of a time series, the partial derivatives of \emph{F} can be similarly simplified.

To compute the total derivative $\dif F / \dif \boldsymbol{x}$ for gradient-based optimization, the chain rule is applied in the following manner:
\begin{equation}
\label{eq:f_total_deriv}
\underbrace{\dfrac{\dif F}{\dif \boldsymbol{x}}}_{1 \times n_x}=\underbrace{\dfrac{\partial F}{\partial \boldsymbol{x}}}_{1 \times n_x} + \underbrace{\dfrac{\partial F}{\partial \boldsymbol{w}}}_{1 \times Kn_w} \underbrace{\dfrac{\dif \boldsymbol{w}}{\dif \boldsymbol{x}}}_{Kn_w \times n_x} ,
\end{equation}
where the partial derivatives $\partial F/\partial \boldsymbol{x}$ and $\partial F/\partial \boldsymbol{w}$ are relatively cheap to compute because they only entail explicit computations.
In contrast, evaluating the total derivative $\dif \boldsymbol{w} /\dif \boldsymbol{x}$ matrix is computationally expensive because $\boldsymbol{w}$ and $\boldsymbol{x}$ are implicitly connected through the residual equations $\boldsymbol{R}(\boldsymbol{w},\boldsymbol{x})=0$.

To compute $\dif \boldsymbol{w} /\dif \boldsymbol{x}$, we can apply the chain rule for $\boldsymbol{R}$. 
Given that the residual equations should always hold, irrespective of the values of design variables $\boldsymbol{x}$.
Therefore, the total derivative $\dif \boldsymbol{R}/\dif \boldsymbol{x}$ must be zero:
\begin{equation}
\label{eq:R_total_deriv}
\dfrac{\dif \boldsymbol{R}}{\dif \boldsymbol{x}}=\dfrac{\partial \boldsymbol{R}}{\partial \boldsymbol{x}} + \dfrac{\partial \boldsymbol{R}}{\partial \boldsymbol{w}} \dfrac{\dif \boldsymbol{w}}{\dif \boldsymbol{x}} = 0 .
\end{equation}
Rearranging the above equation, the following linear system is derived:
\begin{equation}
  \label{eq:dwdx_linear_equation}
  \underbrace{\dfrac{\partial \boldsymbol{R}}{\partial \boldsymbol{w}}}_{Kn_w \times Kn_w} \cdot \underbrace{\dfrac{\dif \boldsymbol{w}}{\dif \boldsymbol{x}}}_{Kn_w \times n_x} = -\underbrace{\dfrac{\partial \boldsymbol{R}}{\partial \boldsymbol{x}}}_{Kn_w \times n_x} .
\end{equation}

Then substituting the solution for $\dif \boldsymbol{w} / \dif \boldsymbol{x}$ from Eq.~(\ref{eq:dwdx_linear_equation}) into Eq.~(\ref{eq:f_total_deriv}) provides:
\begin{equation}
\label{eq:f_total_deriv2}
\underbrace{\dfrac{\dif F}{\dif \boldsymbol{x}}}_{1 \times n_x } = \underbrace{\dfrac{\partial F}{\partial \boldsymbol{x}}}_{1 \times n_x} - \overbrace{\underbrace{\dfrac{\partial F}{\partial \boldsymbol{w}}}_{1 \times Kn_w} \underbrace{ \dfrac{\partial \boldsymbol{R}}{\partial \boldsymbol{w}}^{-1}}_{Kn_w \times Kn_w}}^{\boldsymbol{\psi}^T} \underbrace{\dfrac{\partial \boldsymbol{R} }{\partial \boldsymbol{x}}}_{Kn_w \times n_x} .
\end{equation}

Now we can transpose the Jacobian and solve with $[\partial F / \partial \boldsymbol{w}]^T$ as the right-hand side, which yields the \emph{adjoint equation},
\begin{equation}
  \label{eq:adj_linear_eqn}
  \underbrace{\dfrac{\partial \boldsymbol{R}}{\partial \boldsymbol{w}}^T}_{Kn_w \times Kn_w} \cdot \underbrace{\vphantom{\dfrac{nan}{nan}} \boldsymbol{\psi}}_{Kn_w \times 1} = \underbrace{\dfrac{\partial F}{\partial \boldsymbol{w}}^T}_{Kn_w \times 1} , 
\end{equation}
where $\boldsymbol{\psi}$ is the \emph{adjoint vector}.
Subsequently, the total derivative can be computed by substituting the adjoint vector into Eq.~(\ref{eq:f_total_deriv2}):
\begin{equation}
  \label{eq:adj_total_sens}
  \dfrac{\dif F}{\dif \boldsymbol{x}} = \dfrac{\partial F}{\partial \boldsymbol{x}} - \boldsymbol{\psi}^T \dfrac{\partial \boldsymbol{R}}{\partial \boldsymbol{x}} .
\end{equation}

Since the design variables are not explicitly present in Eq.~\eqref{eq:adj_linear_eqn}, we only need to solve the adjoint equation once for each objective function.
Therefore, the computational cost is independent of the number of design variables but is instead proportional to the number of objective functions.
This method of computing derivatives, as introduced, is also known as the \emph{adjoint method}. 
It is beneficial for field inversion since only one objective function is typically present, even though thousands of design variables may be employed.

The adjoint equation given in Eq.~(\ref{eq:adj_linear_eqn}) can be simplified for the time-marching primal problem. 
As indicated in Eq.~(\ref{eqn_xw_to_r_time_series}), for each $1 \leq i \leq K$, $\boldsymbol{R}^i$  has dependency only on $\boldsymbol{x}$, $\boldsymbol{w}^i$,  $\boldsymbol{w}^{i-1}$, and $\boldsymbol{w}^{i-2}$.
Together with the simplification in Eq.~(\ref{eq:partial_dfdw_simplify}), Eq.~(\ref{eq:adj_linear_eqn}) can be rewritten as:
\begin{equation}
\label{eq:adj_linear_eqn_banded}
\begin{bmatrix}
\frac{\partial \boldsymbol{R}^1}{\partial \boldsymbol{w}^1}^T & \frac{\partial \boldsymbol{R}^2}{\partial \boldsymbol{w}^1}^T & \frac{\partial \boldsymbol{R}^3}{\partial \boldsymbol{w}^1}^T\\
 & \frac{\partial \boldsymbol{R}^2}{\partial \boldsymbol{w}^2}^T & \frac{\partial \boldsymbol{R}^3}{\partial \boldsymbol{w}^2}^T & \frac{\partial \boldsymbol{R}^4}{\partial \boldsymbol{w}^2}^T \\
 &  & \ddots & \ddots \\
 &  & & \frac{\partial \boldsymbol{R}^{K-1}}{\partial \boldsymbol{w}^{K-1}}^T & \frac{\partial \boldsymbol{R}^{K}}{\partial \boldsymbol{w}^{K-1}}^T\\
  &  & & & \frac{\partial \boldsymbol{R}^K}{\partial \boldsymbol{w}^K}^T\\
\end{bmatrix}
\begin{bmatrix}
 \boldsymbol{\psi}^1 \\
 \boldsymbol{\psi}^2 \\
 \vdots \\
 \boldsymbol{\psi}^{K-1} \\
 \boldsymbol{\psi}^K \\
\end{bmatrix}
=
\dfrac{1}{K}
\begin{bmatrix}
 {\frac{\partial f^1 }{\partial \boldsymbol{w}^1}}^T \\
 {\frac{\partial f^2 }{\partial \boldsymbol{w}^2}}^T \\
 \vdots \\
 {\frac{\partial f^{K-1} }{\partial \boldsymbol{w}^{K-1}}}^T \\
 {\frac{\partial f^K }{\partial \boldsymbol{w}^K}}^T \\
\end{bmatrix},
\end{equation}
where the adjoint vector $\boldsymbol{\psi}  \in \mathbb{R}^{Kn_w}$ is broken down into $K$ parts that correspond to the time steps, i.e., $\boldsymbol{\psi}^1, \boldsymbol{\psi}^2, \ldots, \boldsymbol{\psi}^K \in \mathbb{R}^{n_w}$.
Then, Eq.~(\ref{eq:adj_linear_eqn_banded}) can be solved sequentially in a backward manner:
\begin{equation}
  \label{eq:adj_linear_eqn_breakdown}
  \begin{aligned}
  & \dfrac{\partial \boldsymbol{R}^K}{\partial \boldsymbol{w}^K}^T \cdot \boldsymbol{\psi}^K = 
  \dfrac{1}{K}
  \dfrac{\partial f^K}{\partial \boldsymbol{w}^K}^T , \\ 
  & \dfrac{\partial \boldsymbol{R}^{K-1}}{\partial \boldsymbol{w}^{K-1}}^T \cdot \boldsymbol{\psi}^{K-1} = 
  \dfrac{1}{K}
  \dfrac{\partial f^{K-1}}{\partial \boldsymbol{w}^{K-1}}^T 
   - \dfrac{\partial \boldsymbol{R}^K}{\partial \boldsymbol{w}^{K-1}}^T \cdot \boldsymbol{\psi}^K , \\
 & \dfrac{\partial \boldsymbol{R}^i}{\partial \boldsymbol{w}^i}^T \cdot \boldsymbol{\psi}^i = 
 \dfrac{1}{K}
  \dfrac{\partial f^i}{\partial \boldsymbol{w}^i}^T 
   - \dfrac{\partial \boldsymbol{R}^{i+1}}{\partial \boldsymbol{w}^i}^T \cdot \boldsymbol{\psi}^{i+1} 
   - \dfrac{\partial \boldsymbol{R}^{i+2}}{\partial \boldsymbol{w}^i}^T \cdot \boldsymbol{\psi}^{i+2}, \
  K-2 \geq i \geq 1 , 
  \end{aligned}
\end{equation}
which effectively breaks down the original adjoint equation Eq.~(\ref{eq:adj_linear_eqn}) into $K$ much smaller sub-equations.
The right-hand side terms in Eq.~(\ref{eq:adj_linear_eqn_breakdown}) can be efficiently computed using reverse-mode automatic differentiation (AD).
In particular, the matrix-transpose-vector product $[\partial \boldsymbol{R}^{i+2} / \partial \boldsymbol{w}^i]^T \boldsymbol{\psi}^{i+2}$ is evaluated in a Jacobian-free manner immediately after solving for $\boldsymbol{\psi}^{i+2}$,
and $[\partial \boldsymbol{R}^{i+1} / \partial \boldsymbol{w}^i]^T \boldsymbol{\psi}^{i+1}$ is evaluated in a similar manner, then they are passed to the right-hand side of the sub-equation for $\boldsymbol{\psi}^i$.
Note that we solve the adjoint equation~\eqref{eq:adj_linear_eqn_breakdown} in a reverse mode, and computing the matrix-vector products requires access to state variables for all time steps. 
However, storing all state variables in memory is excessively costly. 
Therefore, we write state variables to the disk for all time steps during the primal simulation. 
Then, during the adjoint computation, we read the state variables for each time step from the disk. 

The assumption that the objective function \emph{F} is of the average type in Eq.~(\ref{eq:obj_func}) also simplifies the expression of the total derivative $\dif F / \dif \boldsymbol{x}$ in Eq.~(\ref{eq:adj_total_sens}) as:
\begin{equation}
  \label{eq:adj_total_sens_time_series}
  \dfrac{\dif F}{\dif \boldsymbol{x}} = \sum_{i=1}^K(\dfrac{1}{K} \dfrac{\partial f_i}{\partial \boldsymbol{x}} - {\boldsymbol{\psi}^i}^T \dfrac{\partial \boldsymbol{R}^i}{\partial \boldsymbol{x}}) .
\end{equation}
Hence, we can calculate the total derivative accumulatively as we sequentially solve the sub-equations in Eq.~(\ref{eq:adj_linear_eqn_breakdown}).
This on-the-fly computation of total derivatives is advantageous as it obviates the need to store adjoint vectors for all time steps.

Finally, we elaborate on how to effectively solve the adjoint linear equations in Eq.~(\ref{eq:adj_linear_eqn_breakdown}).
We use the generalized minimal residual (GMRES) solver from the Portable, Extensible Toolkit for Scientific Computation (PETSc) \cite{balay2009petsc} library to solve the adjoint linear equations. 
The GMRES solver achieves quadratic convergence and is notably faster than the fixed-point adjoint solver (linear convergence) employed in our previous studies~\cite{fang2022consistent,fang2024dualitypreserving,fang2024segregated}.
In addition, the GMRES solver's convergence does not require the linear iteration matrix's eigenvalues to be within the unit circle. 
This renders it more robust in practical scenarios, especially when we need to solve the adjoint equations repeatedly. 
For the details implementation of GMRES, one can refer to the study of \citet{fang2024field}.


\section{Results and Discussion}
\label{sec_results}

In this section, we consider the spatial-temporal evolution of unsteady flow over an NACA0012 airfoil undergoing dynamic stall. 
We first run the unsteady CFD simulation and compare the simulated flow fields between the baseline SA model and the reference $k-\omega$ SST model.
We then compare two FIML training strategies for augmenting the SA model toward the reference SST predictions: 1) Steady-state training, which uses lift coefficient ($C_l$) data obtained from steady simulations at multiple angles of attack, and 2) Unsteady training, which uses a time series of drag coefficient ($C_d$) data from unsteady simulations.
We will evaluate the prediction accuracy of the steady- and unsteady-FIML models for drag, lift, moment, surface pressure, and velocity fields.

\subsection{CFD configurations for steady-state and dynamic stall flow simulations for NACA0012}

\begin{figure*}[!t] 
  \centering
  \includegraphics[width=0.8\linewidth]{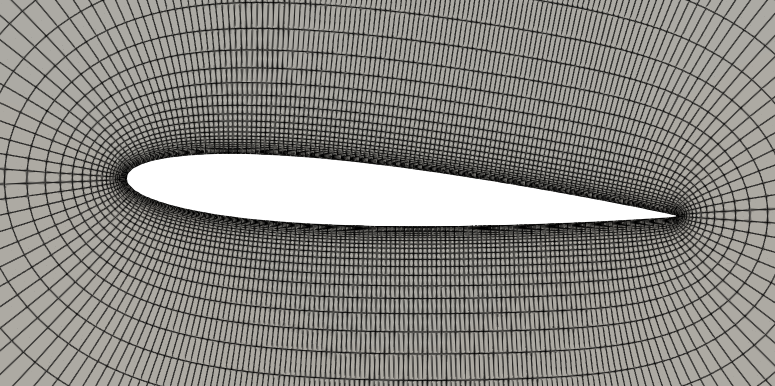}
  \caption{Structured mesh for the unsteady flow over a NACA0012 airfoil with 12,060 cells (initial pitch: $4^{\circ}$).}
  \label{fig_airfoil_mesh}
\end{figure*}

We use the NACA0012 airfoil as the benchmark for the steady-state and the dynamic stall simulations, and the airfoil has a chord length $c=1.0$ m, and starts at a pitch or angle of attack ($\alpha$) of $4^{\circ}$ as shown in Fig.~\ref{fig_airfoil_mesh}.
We use the open-source mesh generation tool \texttt{pyHyp}~\cite{secco2021efficient} to generate the structured mesh, and \texttt{pyHyp} uses hyperbolic volume mesh marching schemes to extrude structured surface meshes into volume meshes.
The computational domain is defined as a cylindrical volume with a radius of $20c$ and a height of $0.1c$.
The leading edge of the airfoil is $20c$ away from the far field boundary.
Finally, we obtain a structured mesh for the airfoil with 12,060 cells.
The computational simulation includes a non-slip boundary condition for the static airfoil and moving wall velocity boundary condition for the dynamic stall case; a velocity inflow and pressure outflow far-field boundary condition; and two symmetry boundaries corresponding to the front and back planes of the simulation domain.
The inlet flow velocity in the far field domain is set as $u_{in}$ = 10 m/s, and the corresponding Reynolds number is $5 \times 10^6$. 

We consider only the pitching-up phase of the airfoil dynamic stall process. The pitching motion of the airfoil starts at an angle of attack of $4^{\circ}$, then a ramp function is used to smoothly increase the pitch rate from 0 to the prescribed value. 
After $t > t_0$, the pitch rate will be held constant until the angle of attack goes past the lift-stall point, just as described in the work of \citet{sharma2019numerical}.
The pitching axis is located at the quarter-chord point of the airfoil, and the ramp function is given by

\begin{equation}
  \label{eq:pitch_rate}
  \Omega^{+}(t) = \Omega^{+}_{0} \left( \frac{\tanh(s(2(t/t_0) - 1)) + \tanh(s)}{1 + \tanh(s)} \right)
\end{equation}
where $\Omega^{+}_{0} = \Omega_0 c / u_{\infty}$, $c$ and $u_{\infty}$ represent the chord length and freestream velocity, respectively.
The parameter $s$ is a scaling parameter that determines the steepness of the ramp function and is set to 2.
A value of 0.35 is used for $t_0$ in the ramp function, and the non-dimensional pitch rate $\Omega^{+}(t)$ is therefore smoothly transitioned from zero at $t = 0$ to $\Omega^{+}_0$ for $t > t_0$.

\begin{table}[!t]
\centering
\caption{Optimization formulation for the steady field inversion problem.}
\begin{tabular}{ l l l r}
\hline
& Function/Variable & Description & Quantity \\
\hline
Min &  $F_1$ & CFD prediction error along with regularization & 1 \\
w.r.t. & $\beta$ & Spatial $\beta$ field for the flow domain & 12,060\\
\hline
\end{tabular}
\label{table_opt_formulation-steady}
\end{table}

\subsection{Steady-FIML using steady-state flow data at different angles of attack}

To avoid the computationally expensive unsteady adjoint solution, previous studies~\cite{fidkowski2022gradient,ahmed2025data} have explored the use of steady-state flow data to augment RANS turbulence models, and then directly apply the augmented model to predict unsteady flows. 
To evaluate the effectiveness of this FIML strategy, we first conduct field inversion across multiple angles of attack (i.e., $\alpha$ = $10^{\circ}$, $12^{\circ}$, $14^{\circ}$, $16^{\circ}$, and $18^{\circ}$) at $Re = 5 \times 10^6$.

Table~\ref{table_opt_formulation-steady} summarizes the field inversion formulation. The objective function $F_1$ is formulated as the weighted sum between the $C_l$ prediction error and a regularization term: 

\begin{equation}
\label{eqn_obj_func_steady}
F_1 = c_1  (C_l^\textrm{CFD} - C_l^\textrm{ref}) ^2 + \frac{c_2}{N_i} \sum_{i=1:N_i}  (\beta_{i} - 1) ^2
\end{equation}
where the subscript $i$ is the mesh cell index, with $N_i$ being the total number of mesh cells, and $\beta$ is the augmentation scalar field to the SA model's production term in Eq.~\eqref{eq_sa_model},
$c_1 = 1.0$ and $c_2 = 0.01$ are the weights for the two terms in the objective; $c_2$ is mainly used to control the regularization.
The design variables are the local $\beta$ values defined at each cell in the flow field. With 12,060 cells in total, there are 12,060 design variables.
Note that in this study, the $C_l$ predictions from the SST turbulence model are used as the reference to demonstrate the proposed FIML frameworks.
In future work, we will use eddy-resolving models, such as large-eddy simulation (LES), as the reference to enhance the traiend model’s fidelity and practical applicability.

\begin{figure*}[!t]
  \centering
  \includegraphics[width=0.48\linewidth]{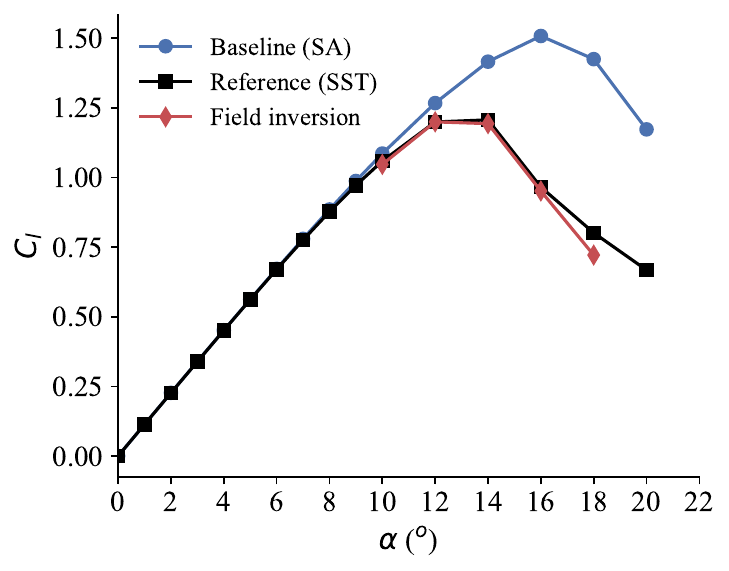}
  \caption{Steady-state field inversion for the NACA0012 airfoil at angles of attack $\alpha$ = $10^{\circ}$, $12^{\circ}$, $14^{\circ}$, $16^{\circ}$, and $18^{\circ}$.}
  \label{fig:training-steady}
\end{figure*}

Fig.~\ref{fig:training-steady} illustrates the comparison of the lift coefficient ($C_l$) versus angle of attack ($\alpha$) curve.
In the linear region ($\alpha \approx 0^\circ$ to $10^\circ$), where the flow is typically laminar and attached, the baseline (SA) and the reference (SST) show excellent agreement with each other.
This indicates that for attached flow conditions, there is little room to improve the turbulence model.
During the onset of the separation and stall region ($\alpha \approx 12^\circ$ to $14^\circ$), the baseline model significantly over-predicts the maximum lift coefficient, compared with the reference SST model.
Moreover, it also delays the stall (the angle at which $C_l$ peaks) to $\alpha \approx 16^\circ$. 
The field inversion achieves a much closer agreement with the reference data.
Both field inversion and the reference model have approximately the same maximum lift coefficient and the stall angle ($\alpha \approx 14^\circ$).
Field inversion successfully corrects the overall prediction of lift and the stall delay exhibited by the baseline model, making its prediction nearly identical to the reference data.
In the post-stall regime ($\alpha > 14^\circ$), the baseline model continues to diverge, maintaining a significantly higher lift coefficient than the reference data.
However, the field inversion model tracks the sharp drop in lift and generally follows the trend of the reference data. 
Although the field inversion curve shows a slightly steeper lift decay compared to the reference data after $\alpha \approx 16^\circ$, it remains significantly closer to the reference data than the baseline model.

\begin{figure*}[!t]
  \centering
  \includegraphics[width=0.48\linewidth]{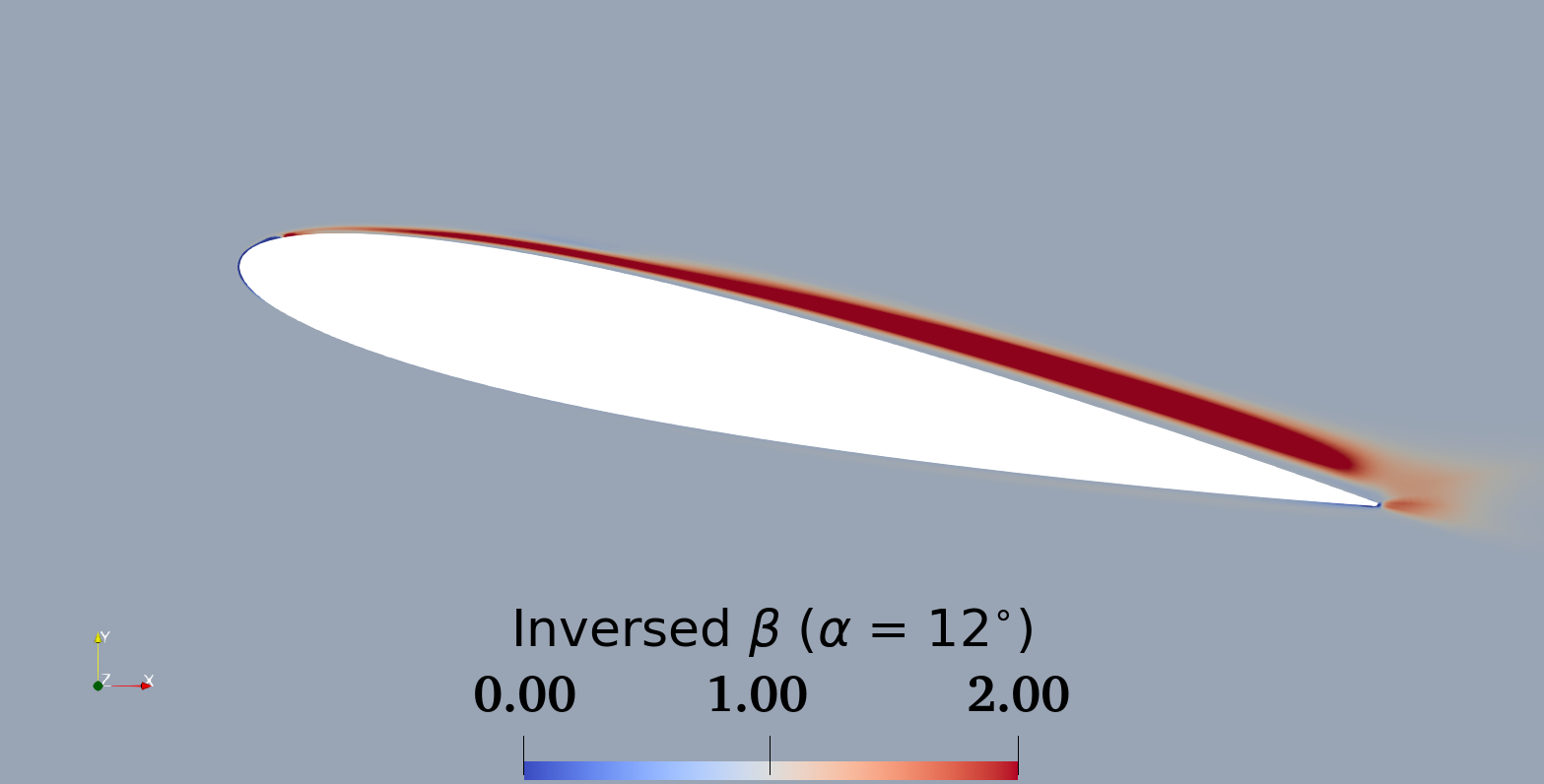}
  \includegraphics[width=0.48\linewidth]{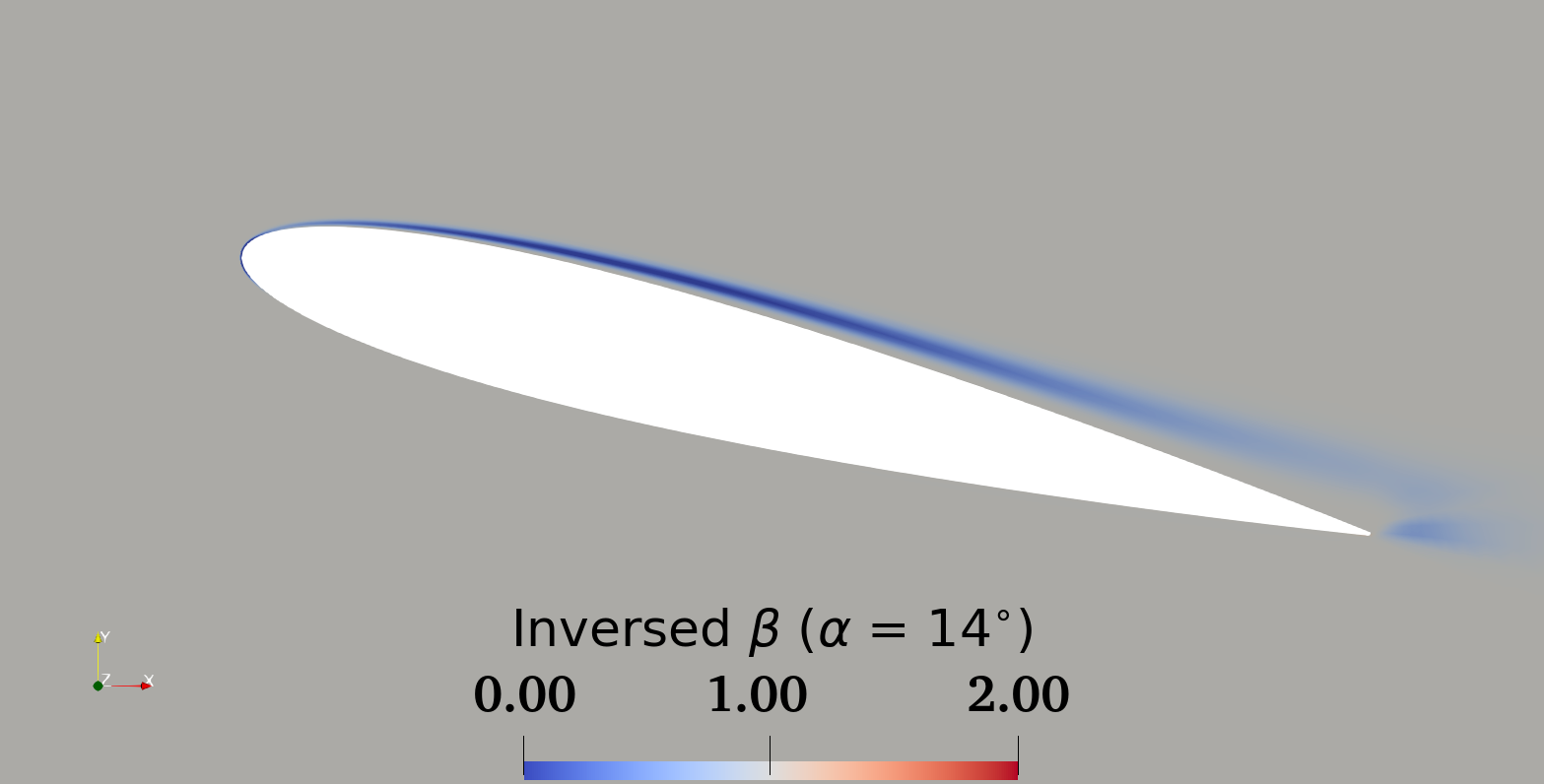} \\
  \includegraphics[width=0.48\linewidth]{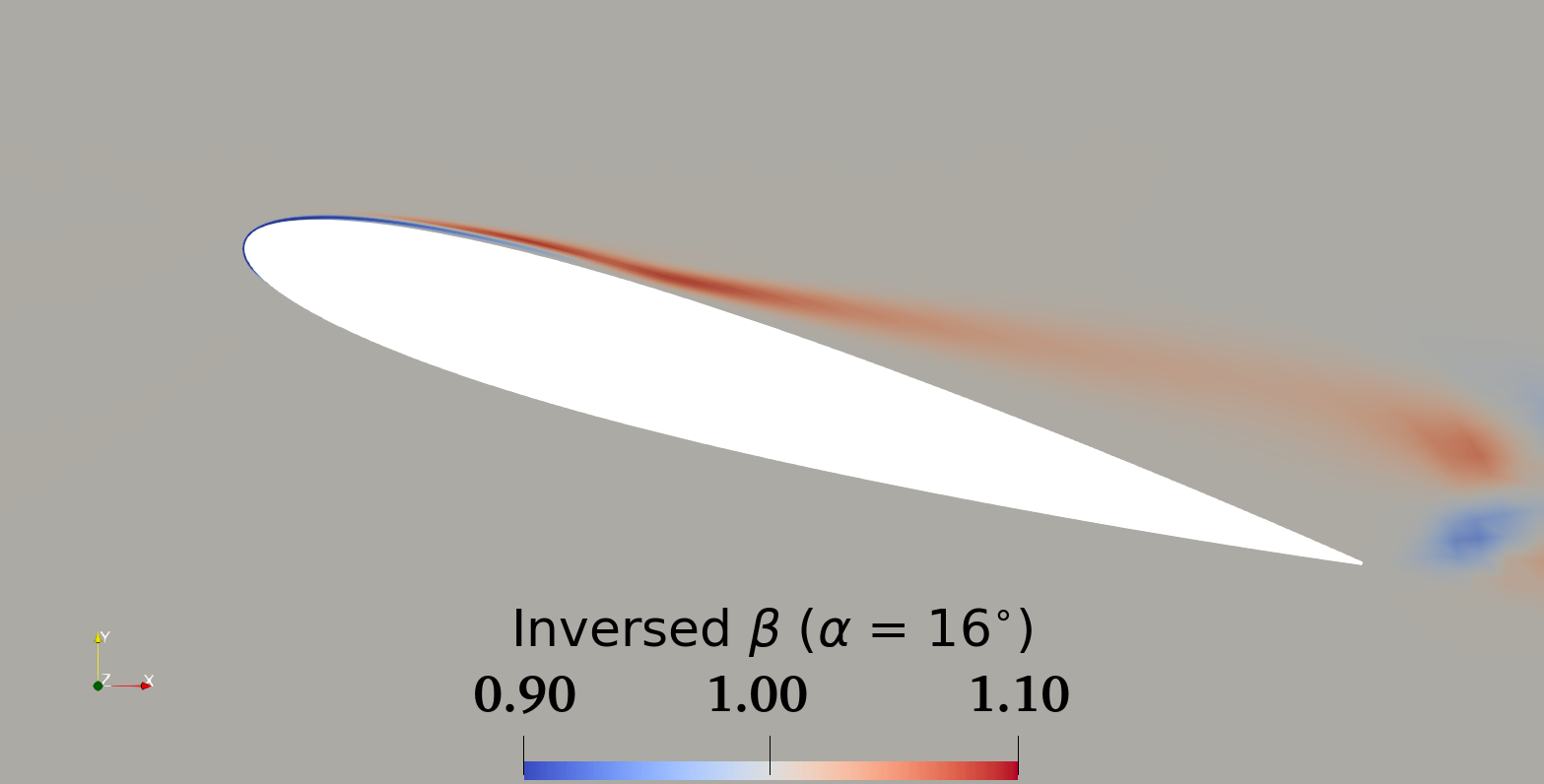}
  \includegraphics[width=0.48\linewidth]{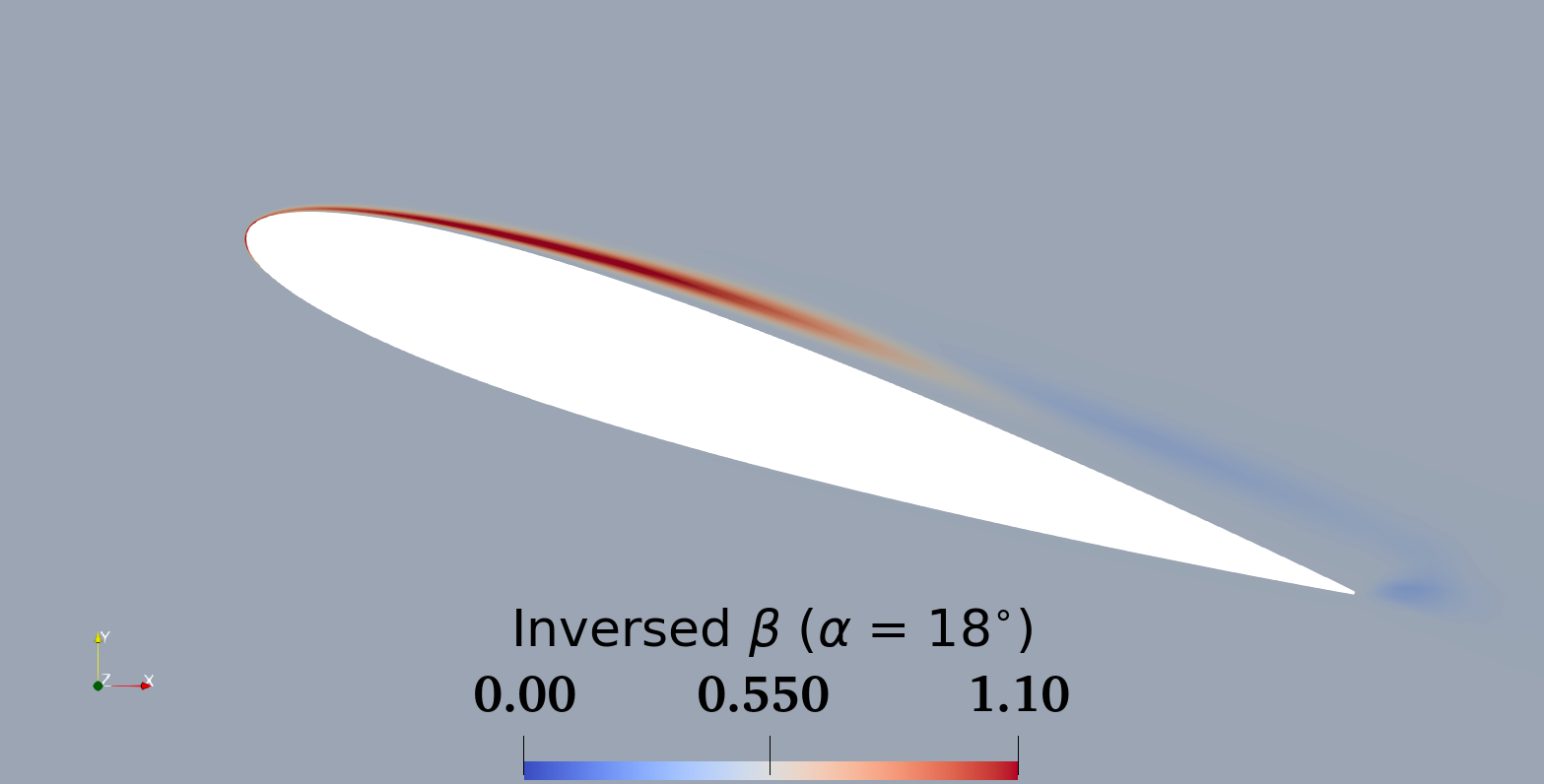}
  \caption{Inverse $\beta$ field for NACA0012 airfoil across angles of attack $\alpha$ = $12^{\circ}$, $14^{\circ}$, $16^{\circ}$, and $18^{\circ}$.}
  \label{fig:beta-steady}
\end{figure*}

The inverse $\beta$ fields for the NACA0012 airfoil are shown in Fig.~\ref{fig:beta-steady}.
These figures clearly illustrate the evolution of the model correction as the angle of attack increases, pitching toward and past the static stall angle (which is typically around $12^\circ - 16^\circ$ for the NACA0012 airfoil).
In the onset of the separation regime ( $\alpha = 12^\circ$), a strong and concentrated red region ($\beta > 1.0$) is present very close to the airfoil's suction surface, particularly in the attached boundary layer and the trailing edge shear layer.
This indicates the baseline model requires a significant increase in turbulence production in the boundary layer to match the target reference data.
In the imminent stall stage ($\alpha = 14^\circ$), the $\beta$ field dampens the turbulence production near the airfoil suction surface, particularly in the leading-edge shear layer and along the separated flow region extending downstream. 
In these areas, $\beta < 1.0$, suggesting an over-prediction of turbulence production by the baseline model.
In the post-stall stage ($\alpha = 16^\circ$), the $\beta$ field becomes more complicated, showing a mix of red ($\beta > 1.0$) in the leading-edge shear layer and blue ($\beta < 1.0$) further downstream in the wake.
This suggests that the field inversion is locally adjusting the model behavior in the leading-edge shear layer to enable the baseline model to more accurately capture the flow separation and transition.
At the fully developed stall condition ($\alpha = 18^\circ$), a narrow red region ($\beta > 1.0$) persists in the vicinity of the leading edge, followed downstream by a light blue region ($\beta < 1.0$) within the separated flow and wake. 
This indicates locally enhanced turbulence production near the leading edge, where the shear layer originates, and a gradual attenuation of turbulence production throughout the separated region and wake.

\begin{table}[!t]
\centering
\caption{Four normalized local flow features used in this study.}
\begin{tabular}{ l l l r}
\hline
Feature & Formulation & Description \\
\hline
$\eta_1$ &  $P/(P+D)$ & Ratio of the turbulence production and destruction term  \\
$\eta_2$ &  $|\Omega| / (|\Omega| + |S|) $ & Ratio of the vorticity and strain magnitudes  \\
$\eta_3$ &  $\tilde{\nu} / (\tilde{\nu} + \nu) $ & Ratio of the turbulence and kinematic viscosity \\
$\eta_4$ &  $\sqrt{\frac{\partial{p}}{\partial{x_i}} \frac{\partial{p}}{\partial{x_i}}}  / (\sqrt{\frac{\partial{p}}{\partial{x_i}} \frac{\partial{p}}{\partial{x_i}}} + \frac{\partial{U_k^2}}{\partial{x_k}})$ & Ratio of pressure normal stress to shear stress \\
\hline
\end{tabular}
\label{table_flow_features}
\end{table}


After all the field inversion cases are finished, we train a multi-layer neural network model to connect the optimized $\beta$ fields and local flow features.
A four-layer neural network architecture is employed, consisting of an input layer, two hidden layers, and an output layer. 
The model uses four local flow features as inputs and outputs the corresponding augmentation $\beta$ field. 
For the steady-FIML case, each hidden layer is configured with 20 neurons.
During training, the loss function decreases by approximately three orders of magnitude.
While the selection of flow features has a significant impact on the FIML accuracy and generality, we choose four flow features used in our previous studies~\cite{fang2024field}.
The description of the four selected flow features is summarized in Table~\ref{table_flow_features}.
Note that $P$ and $D$ in flow feature $\eta_1$ are the turbulence production and destructions terms in Eq.~\eqref{eq_sa_model},  where $P = C_{b1} \tilde{S} \tilde{\nu}$, and $D = C_{w1} f_w \left( \dfrac{\tilde{\nu}}{d} \right)^2$.
The $\Omega$ and $S$ in flow feature $\eta_2$ are the rotation and stress tensors, where $|\Omega| = \sqrt{\Omega_{ij} \Omega_{ij}}$, $\Omega_{ij} = \frac{1}{2} \left(\frac{\partial u_i}{\partial x_j} - \frac{\partial u_j}{\partial x_i} \right)$, $|S| = \sqrt{S_{ij} S_{ij}}$, and $S_{ij} = \frac{1}{2} \left(\frac{\partial u_i}{\partial x_j} + \frac{\partial u_j}{\partial x_i} \right)$. 
The trained model will then be used to predict time-resolved unsteady flows during dynamic stall.

\subsection{Unsteady-FIML using time series of drag coefficient}

Table~\ref{table_opt_formulation-unsteady} summarizes the unsteady field inversion formulation. We use the time series of drag coefficient ($C_d$) as the training data; therefore, the objective function $F_2$ is as follows:

\begin{equation}
\label{eqn_obj_func_unsteady}
F_2 = \frac{1}{K}\sum_{t=1:K} [c_1  (C_{d, t}^\textrm{CFD} - C_{d, t}^\textrm{ref}) ^2 + \frac{c_2}{N_i} \sum_{i=1:N_i}  (\beta_{i, t} - 1) ^2]
\end{equation}
where the subscript $t$ denotes the time index, the subscript $i$ is the mesh cell index, $N_i$ is the total number of mesh cells, $K$ is the number of time steps, $\beta$ is also the augmentation scalar field to the SA model's production term in Eq.~\eqref{eq_sa_model}.
To avoid over-fitting, we also add a regularization term to minimize the spatial variations of the augmentation scalar field $\beta$ at all time steps.
$c_1 = 1.0$ and $c_2 = 0.01$ are the weights for the two terms in the objective.
The design variables are the time-evolving $\beta$ fields.
We run the unsteady field inversion for 0.7 s, and the time step $\Delta t = 0.0005$s, so we have a total of 1,400 time steps, directly selecting the $\beta$ field at every time step would lead to an impractically large number of design variables (12,060 $\times$ 1,401 = 16,896,060), making the unsteady FIML optimization computationally infeasible.
To reduce the dimensionality, we instead sample the $\beta$ fields every 70 time steps, resulting in a total number of 21 $\beta$ fields used as design variables; the total number of independent design variables is 12,060 $\times$ 21 = 253,260.
In the CFD simulation, the optimizer assigns values for the sampled $\beta$ fields, and the intermediate $\beta$ values between the sampled $\beta$ fields are linearly interpolated.


\begin{table}[!t]
\centering
\caption{Optimization formulation for the unsteady field inversion problem.}
\begin{tabular}{ l l l r}
\hline
& Function/Variable & Description & Quantity \\
\hline
Min &  $F_2$ & CFD prediction error along with regularization & 1 \\
w.r.t. & $\beta$ & Spatial $\beta$ field for the flow domain & 253,260\\
\hline
\end{tabular}
\label{table_opt_formulation-unsteady}
\end{table}

\begin{figure*}[!t]
  \centering
  \includegraphics[width=0.48\linewidth]{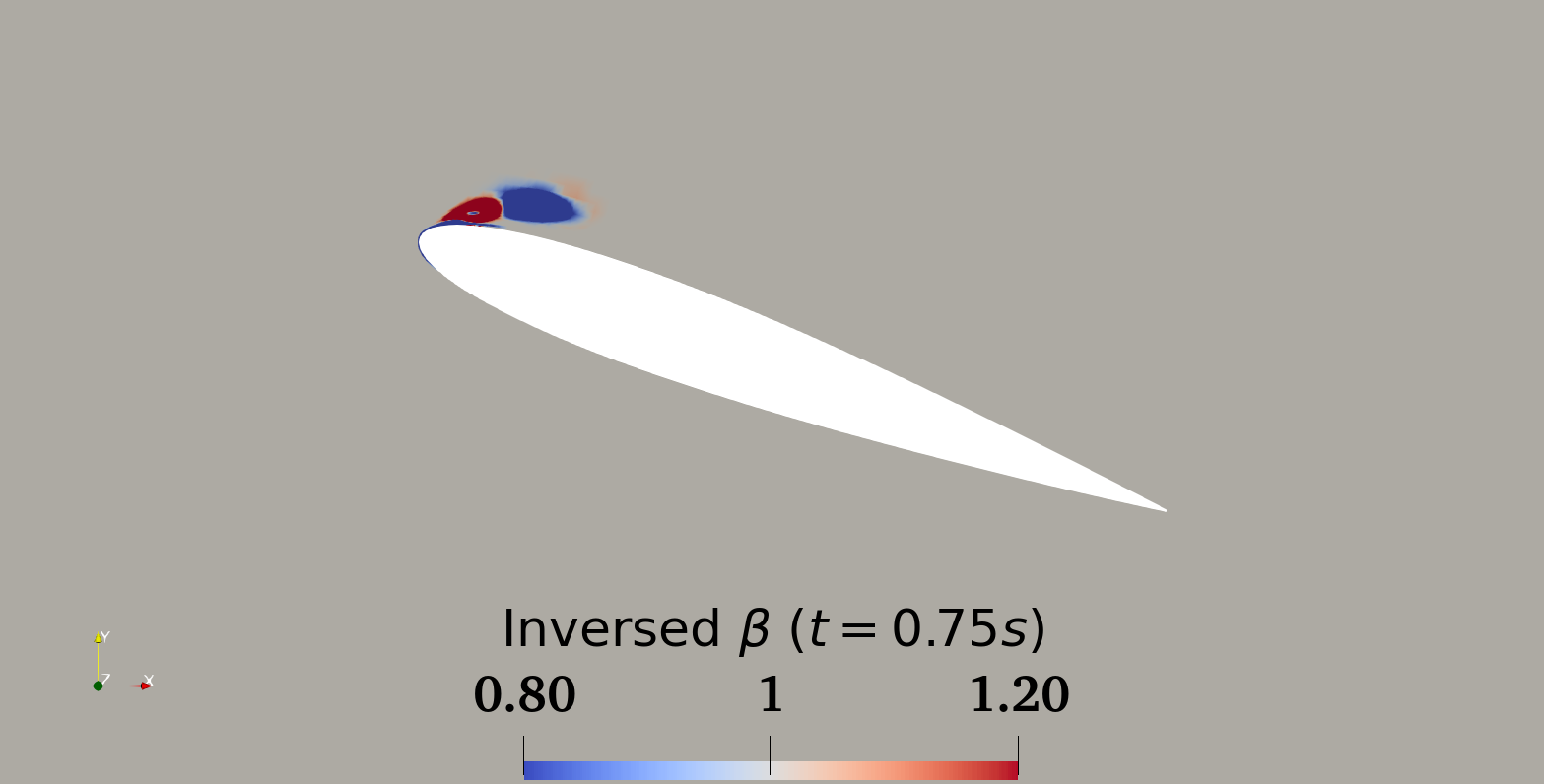}
  \includegraphics[width=0.48\linewidth]{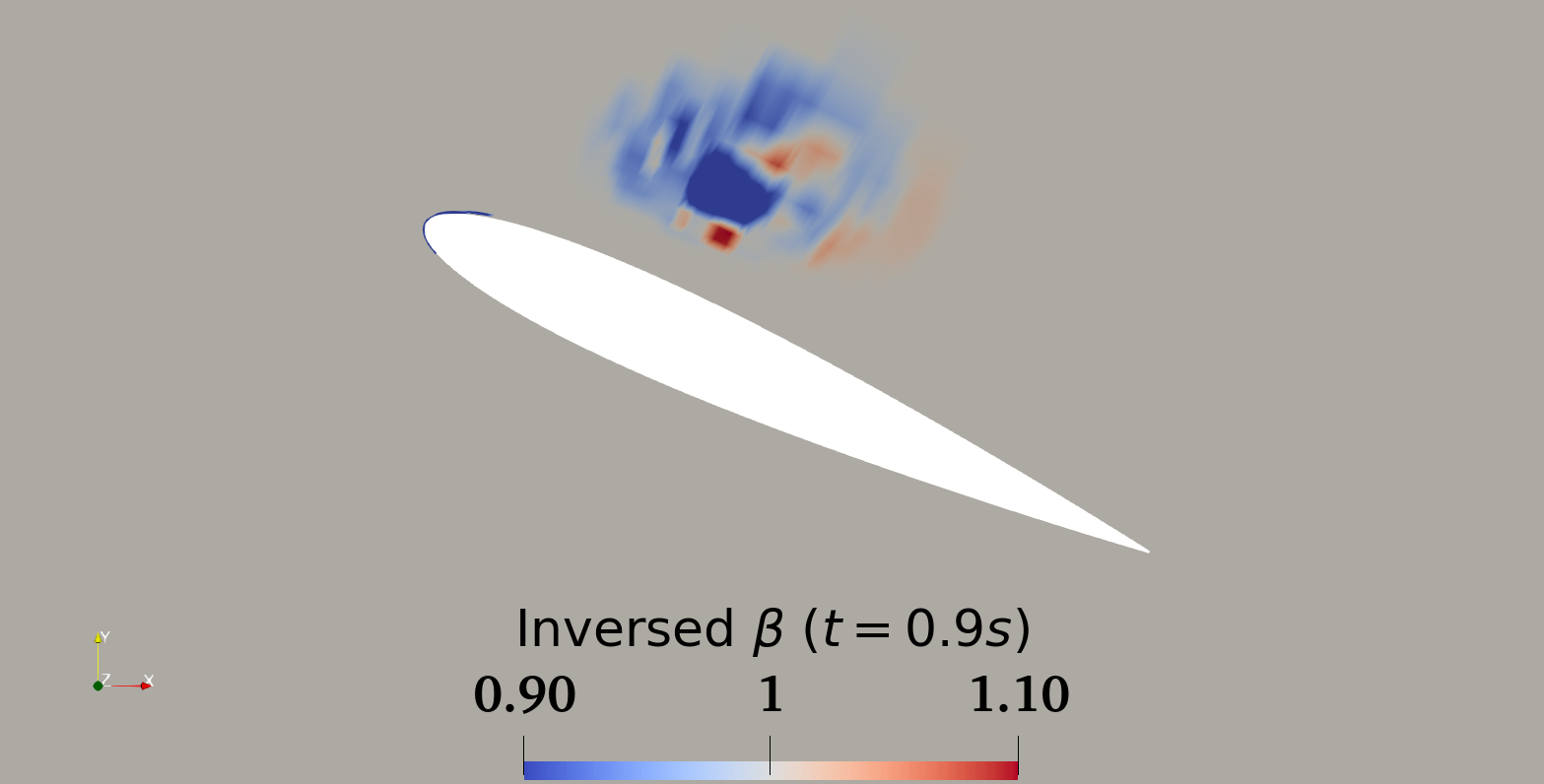}\\
  \includegraphics[width=0.48\linewidth]{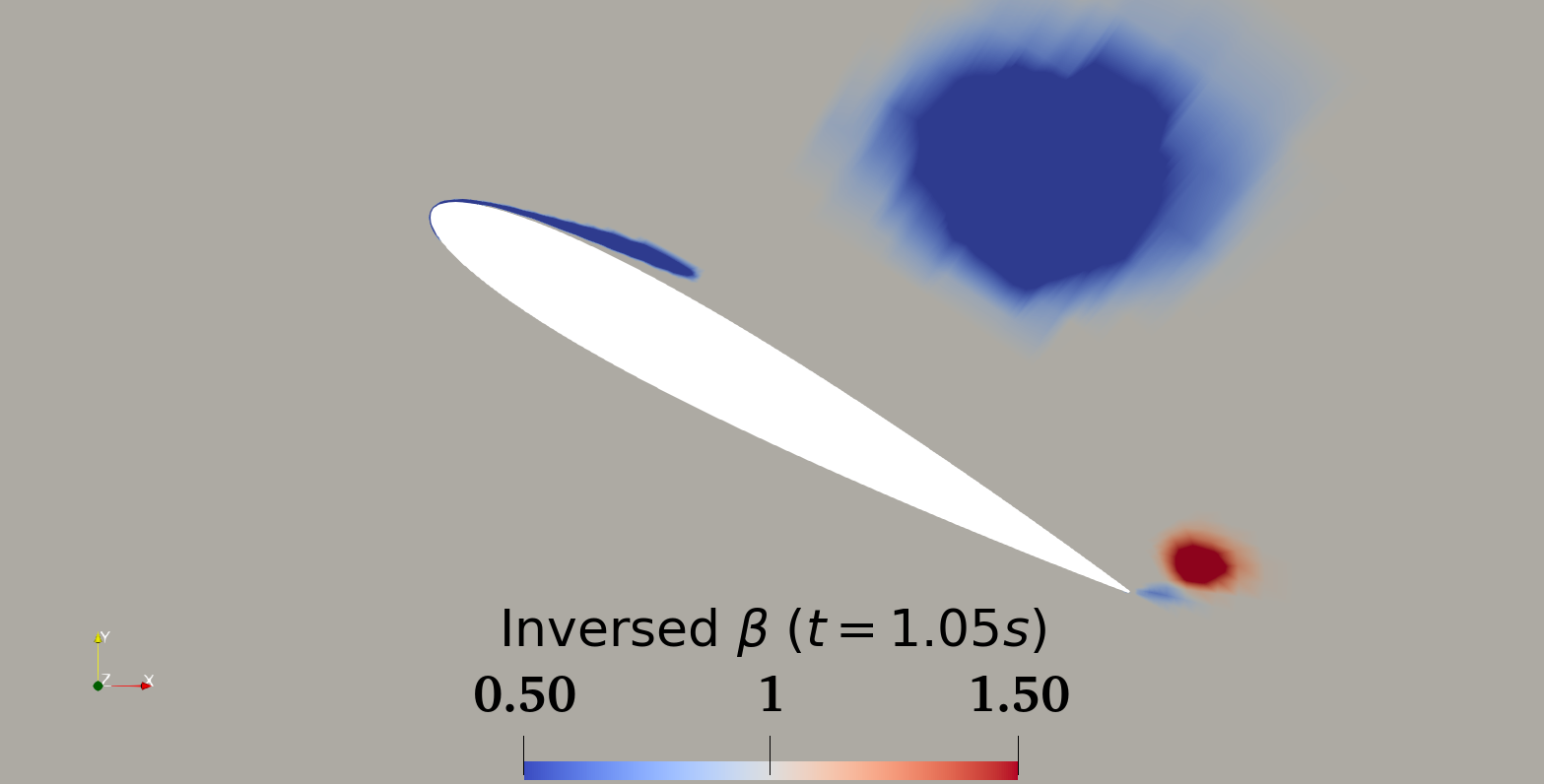}
  \includegraphics[width=0.48\linewidth]{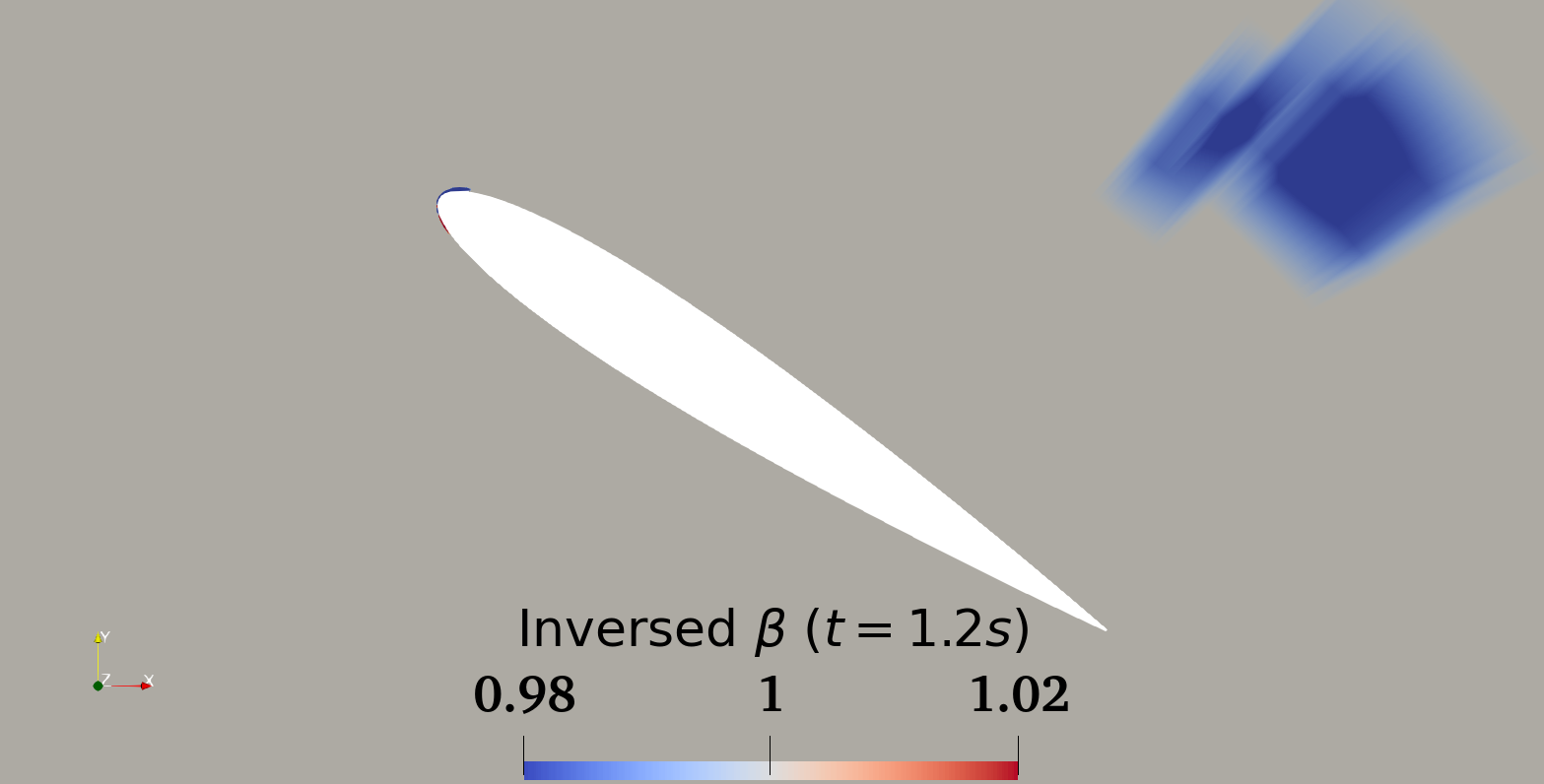}
  \caption{Inverse $\beta$ field at four time instances t=0.75, 0.9, 1.05, and 1.2s, respectively.}
  \label{fig:beta-unsteady}
\end{figure*}

Fig.~\ref{fig:beta-unsteady} shows the contour plots of the inverse $\beta$ fields (the spatial scalar correction applied to the baseline turbulence production term) around the NACA0012 airfoil at four sequential time instances.
The time evolution of the $\beta$ fields illustrates the spatial-temporal nature of the model corrections during the dynamic stall.
At the early stage of flow ($t = 0.75$ s), the $\beta$ field exhibits a spatially localized correction pattern concentrated near the airfoil leading edge.
A narrow region of turbulence enhancement ($\beta > 1$, red contour) is visible near the leading edge of the airfoil, followed by a small region of turbulence damping ($\beta < 1$, blue contour).
By $t = 0.9$ s, the $\beta$ correction field becomes more extended and exhibits a distinct, spatially complex pattern corresponding to the forming and growing dynamic stall vortex.
A mix of intense turbulence damping region (blue contour), coupled with a slightly scattered turbulence enhancement region, appears above the suction surface of the airfoil.
At $t = 1.05$ s, a large and circular strong turbulence damping region occupies the main separated flow area above the airfoil, corresponding to the detached dynamic stall vortex; while a distinct and localized concentrated turbulence enhancement region emerges at the trailing edge of the airfoil, indicating the turbulence activity associated with vortex convection and wake unsteadiness.
Additionally, a distinct thin region of turbulence damping remains attached to the leading edge of the airfoil and spreads along the upper surface. 
By $t = 1.2$ s, the dominant low $\beta$ distribution convects further downstream, and the overall correction magnitude is significantly smaller than previous time instances, indicating that once the intense dynamic stall vortex moves away, the localized turbulence correction requirement rapidly decreases.

\begin{figure*}[!t]
  \centering
  \includegraphics[width=0.48\linewidth]{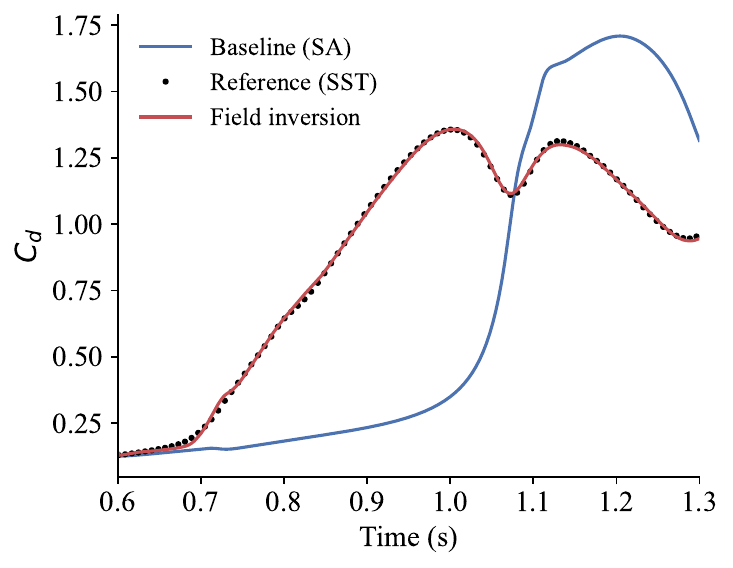}
  \includegraphics[width=0.48\linewidth]{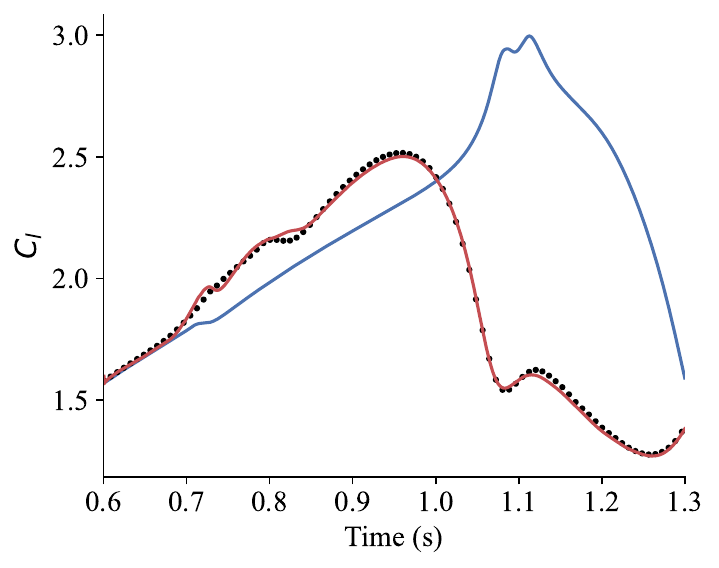} \\
  \includegraphics[width=0.48\linewidth]{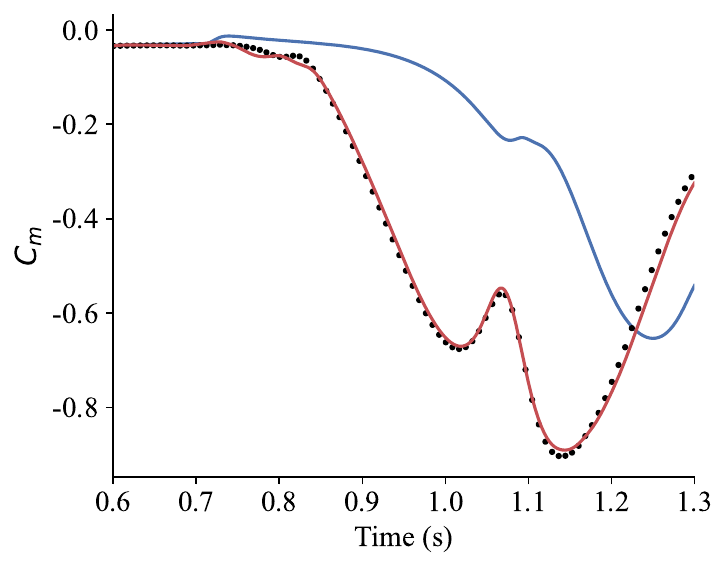}
  \caption{Time series of drag, lift, and pitching moment among baseline, reference, and Field inversion.}
  \label{fig:using-0.5-field-inversion}
\end{figure*}

Fig.~\ref{fig:using-0.5-field-inversion} presents the temporal evolution of drag, lift, and pitching moment for the NACA0012 airfoil under dynamic stall.
The baseline model exhibits substantial discrepancies in all three aerodynamic coefficients, under-prediction of drag during the early pitch-up phase, delayed stall onset, excessive lift overshoot, and noticeable phase lag in the pitching moment.
In contrast, the field inversion model demonstrates excellent agreement; it accurately captures the phase and amplitude of the drag increase, lift rise, stall onset, and moment variation throughout the unsteady dynamic stall.


Similar to the steady-FIML case, we then used the 21 optimized, sampled $\beta$ fields, along with the corresponding local flow features, as the data to train a multi-layer neural network model.
Again, the neural network model has four layers, including an input layer, two hidden layers, and an output layer. 
The network takes four local flow features as inputs and predicts the corresponding augmentation $\beta$ field. 
Due to the increased complexity in the unsteady-FIML case, 100 neurons are used in each hidden layer. 
The loss function drops by about three orders of magnitude during training.
The trained model will then be used to predict the spatial-temporal evolution of flow fields in dynamic stall.

\subsection{Performance Evaluation of FIML Models Trained and Tested at the Same Pitch Rate (0.5 rad/s)}


In this section, we evaluate the performance of steady- and unsteady-FIML models in predicting the unsteady aerodynamics of airfoils undergoing dynamic stall.
Here we briefly discuss the FIML setup, and more details of the training are presented in the Appendix.
For the steady-FIML framework, in the stage of steady field inversion, the reference data can be a scalar (e.g., lift or drag coefficient of an airfoil), surface variable (e.g., skin friction), or field variable (e.g., pressure or velocity field).
In this study, we obtain the optimized $\beta$ field using the lift coefficient from the reference $k-\omega$ SST model. 
The objective function $F_1$ is formulated as: 

\begin{equation}
\label{eqn_argmin_beta}
F_1 = c_1  (C_l^\textrm{CFD} - C_l^\textrm{ref}) ^2 + \frac{c_2}{N_i} \sum_{i=1:N_i}  (\beta_{i} - 1) ^2
\end{equation}
where the subscript $i$ is the mesh cell index, with $N_i$ being the total number of mesh cells, and $\beta$ is the augmentation scalar field to the SA model's production term in Eq.~\eqref{eq_sa_model},
$c_1 = 1.0$ and $c_2 = 0.01$ are the weights for the two terms in the objective, $c_2$ is mainly used to control the regularization.

For the unsteady-FIML framework, we use the time series of drag coefficient ($C_d$) as the training data; therefore, the objective function $F_2$ is as follows:

\begin{equation}
\label{eqn_obj_func}
F_2 = \frac{1}{K}\sum_{t=1:K} [c_1  (C_{d, t}^\textrm{CFD} - C_{d, t}^\textrm{ref}) ^2 + \frac{c_2}{N_i} \sum_{i=1:N_i}  (\beta_{i, t} - 1) ^2]
\end{equation}
where the subscript $t$ denotes the time index, the subscript $i$ is the mesh cell index, $N_i$ being the total number of mesh cells, $K$ is the number of time steps, $\beta$ is also the augmentation scalar field to the SA model's production term in Eq.~\eqref{eq_sa_model}.
To avoid over-fitting, we also add a regularization term to minimize the spatial variations of the augmentation scalar field $\beta$ at all time steps.
$c_1 = 1.0$ and $c_2 = 0.01$ are the weights for the two terms in the objective.

\begin{figure*}[!t]
  \centering
  \includegraphics[width=0.48\linewidth]{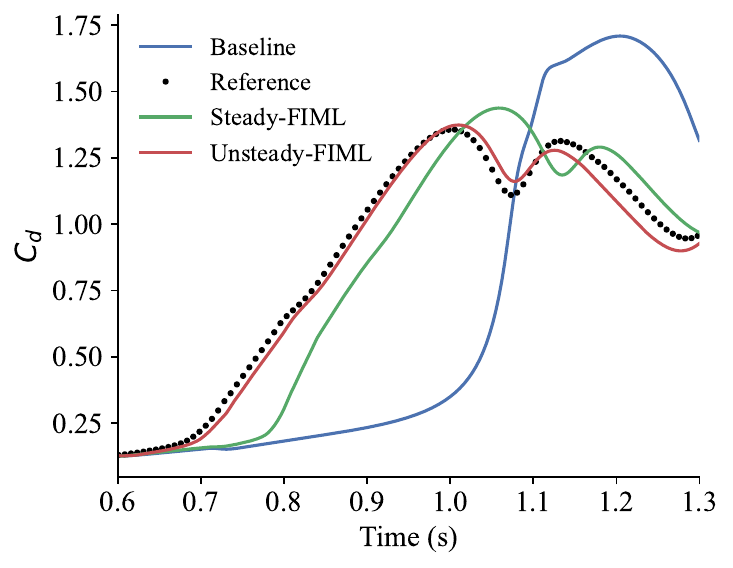}
  \includegraphics[width=0.48\linewidth]{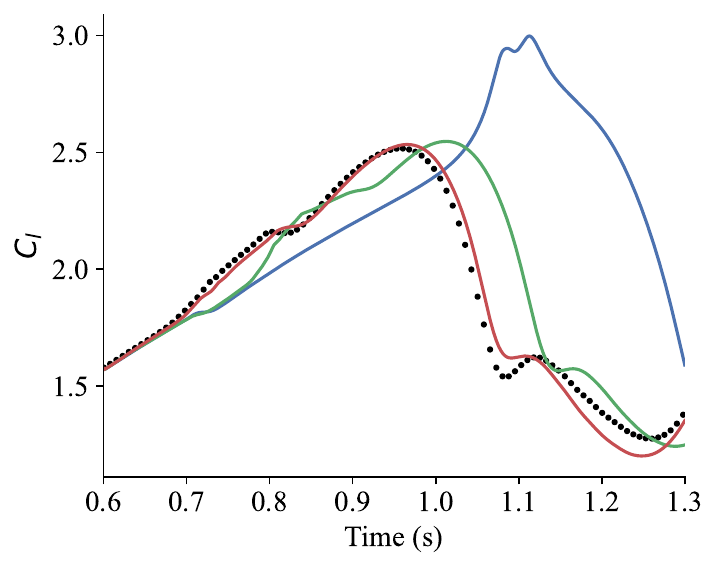} \\
  \includegraphics[width=0.48\linewidth]{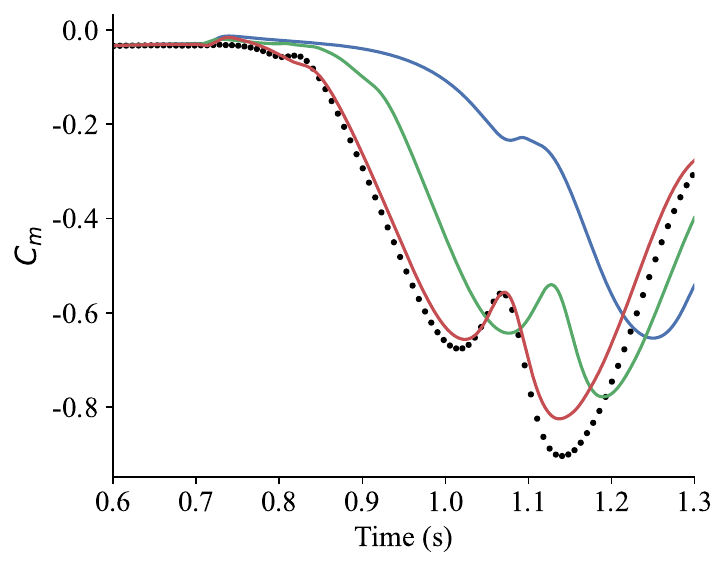}
  \caption{Time series of airfoil drag, lift, and pitching moment among baseline, reference, steady-and unsteady-FIML (pitch rate 0.5 rad/s).}
  \label{fig:cd_cl_cm-steady-unsteady}
\end{figure*}

Fig.~\ref{fig:cd_cl_cm-steady-unsteady} illustrates the temporal evolution of drag, lift, and pitching moment for the pitching NACA0012 airfoil at $Re = 5 \times 10^6$.
For the drag coefficient, the baseline model exhibits significant discrepancies when compared to the reference data and underestimates the drag during the initial pitch-up motion (t $\approx 0.7$ to 1.0 s), followed by a delayed and pronounced peak occurring around t $\approx 1.2$ s.
The steady-FIML model demonstrates an apparent improvement by initiating the drag rise earlier than the baseline, indicating a partial correction of the turbulence model.
However, the steady-FIML model's prediction continues to under-predict the drag magnitude relative to the reference data during the primary drag rise phase and fails to accurately capture the complex transient peak and trough in the dynamic stall process.
As for the lift coefficient, the baseline model demonstrates a substantial over-prediction of the maximum $C_l$ ($\approx 3.0$) and a delayed stall onset occurring approximately $t \approx 1.15$ s.
In contrast, the steady-FIML model effectively mitigates the behavior by reducing the maximum $C_l$ magnitude and advancing the onset of stall, achieving close agreement with the reference data.
However, the steady-FIML model does not fully capture the abrupt nature of lift decay and exhibits a phase shift when compared to the reference data.
As for the pitching moment, the reference data displays characteristic sharp deep decreases (troughs near $t \approx 1.0$ and 1.14 s), while the baseline model entirely misses the phase and magnitude of these critical characteristics. 
The steady-FIML model tracks the overall nose-down trend, while superior to the baseline, but fails to resolve the magnitude and timing of these transient fluctuations.

Although the steady-FIML model (green curves) shows reasonable improvements over the baseline predictions (including better phase alignment and reduced peak over-predictions), notable discrepancies persist throughout the dynamic stall process.
The steady-FIML model captures the general temporal trends and qualitatively reproduces the stall evolution, yet errors remain in both magnitude and phase across all the aerodynamic coefficients.
These results highlight the inherent limitation of using steady-state training data; though the steady-FIML model is trained across a range of angles of attack, it cannot fully capture the physics of inherently unsteady dynamic stall phenomena, where flow-history effects, unsteady separation, and transient vortex dynamics play dominant roles. 
This highlights the benefit of the time-resolved unsteady FIML model in capturing the necessary spatial-temporal evolution of flow fields during dynamic stall.
As shown in Fig.~\ref{fig:cd_cl_cm-steady-unsteady}, the unsteady-FIML (red curves) consistently outperforms the steady-FIML in predicting all key flow physics, such as the maximal lift and timing of stall onset.

\begin{figure*}[!t]
  \centering
  \includegraphics[width=0.48\linewidth]{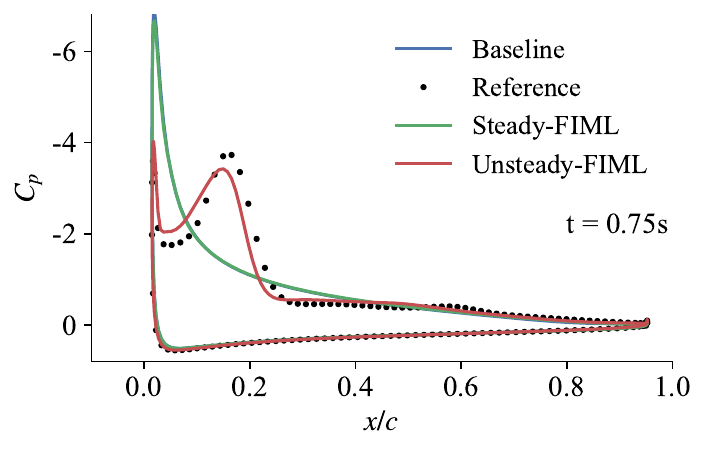}
  \includegraphics[width=0.48\linewidth]{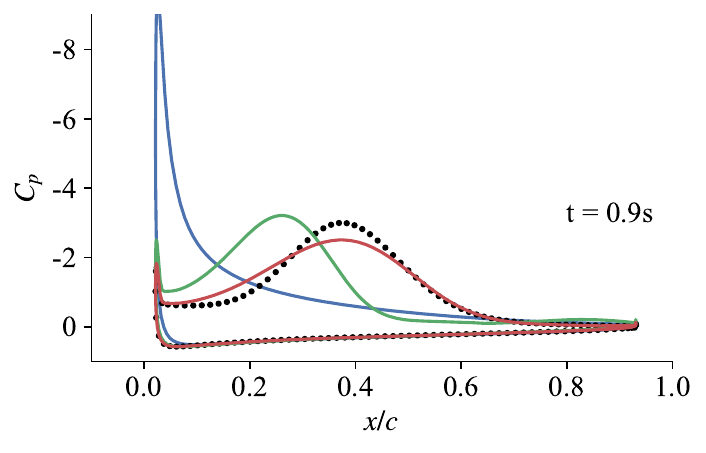} \\
  \includegraphics[width=0.48\linewidth]{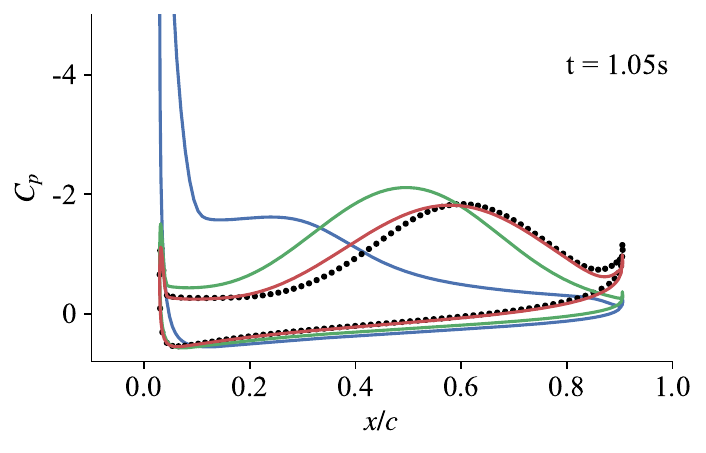}
  \includegraphics[width=0.48\linewidth]{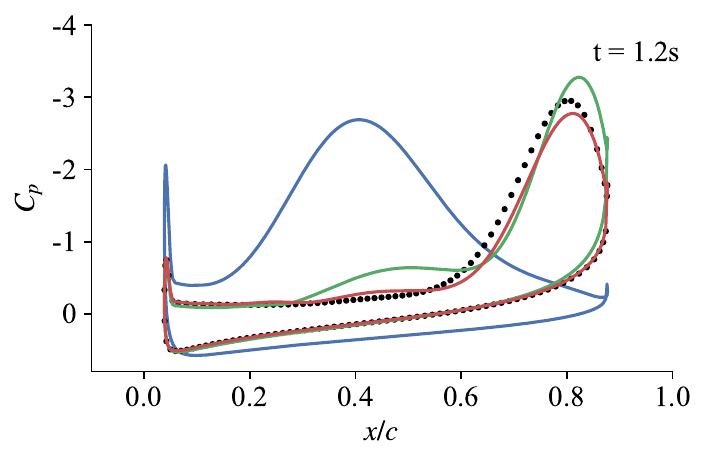}
  \caption{Surface pressure profiles at various time instances among baseline, reference, steady- and unsteady-FIML (pitch rate 0.5 rad/s).}
  \label{fig:cp-steady-unsteady}
\end{figure*}

Fig.~\ref{fig:cp-steady-unsteady} presents the temporal evolution of the surface pressure coefficient distributions at four representative time instances.
At $t = 0.75$ s, both the baseline and steady-FIML models produce nearly identical $C_p$ distributions (the green and blue curves overlap), indicating that the steady-FIML model fails to correct the SA model during the early phase of flow development.
Both models predict a large suction spike ($C_p \approx -7$) near the leading edge, substantially exceeding the reference solution's peak magnitude of $C_p \approx -4$.
And both models fail to capture the suction bump observed in the reference solution over the leading-edge region ($ x/c < 0.3$).
As the flow develops ($t = 0.9$ s), the baseline model continues to predict a severe leading-edge suction peak, further deviating from the reference profile. 
The steady-FIML model, in contrast, achieves reasonable correction by eliminating the leading-edge peak and agrees better with the reference.
Nevertheless, the steady-FIML model's prediction exhibits notable quantitative discrepancies, particularly in the suction region ($ x/c < 0.6$), where it fails to accurately capture the phase of the suction bump.
At $t = 1.05$ and $t=1.2$ s, the baseline model still produces a large leading-edge suction spike with a pressure distribution that has little resemblance to the reference solution.
Whereas the steady-FIML correction removes this peak and reproduces the qualitative trend in the suction side seen in the reference, especially at $t=1.2$.

These instantaneous pressure distributions reveal fundamental limitations of the trained steady-FIML model when applied to highly transient dynamic stall phenomena, though the steady-FIML model demonstrates progressive improvement on the $C_p$ distribution as the flow evolves.
Again, this is primarily attributed to the lack of the temporal information necessary to capture flow history effects in the dynamic stall event.
In contrast, the unsteady FIML agrees well with the reference, including the beginning of the dynamic stall ($t=0.75$ s).

\begin{figure*}[!t]
  \centering
  \includegraphics[width=0.6\linewidth]{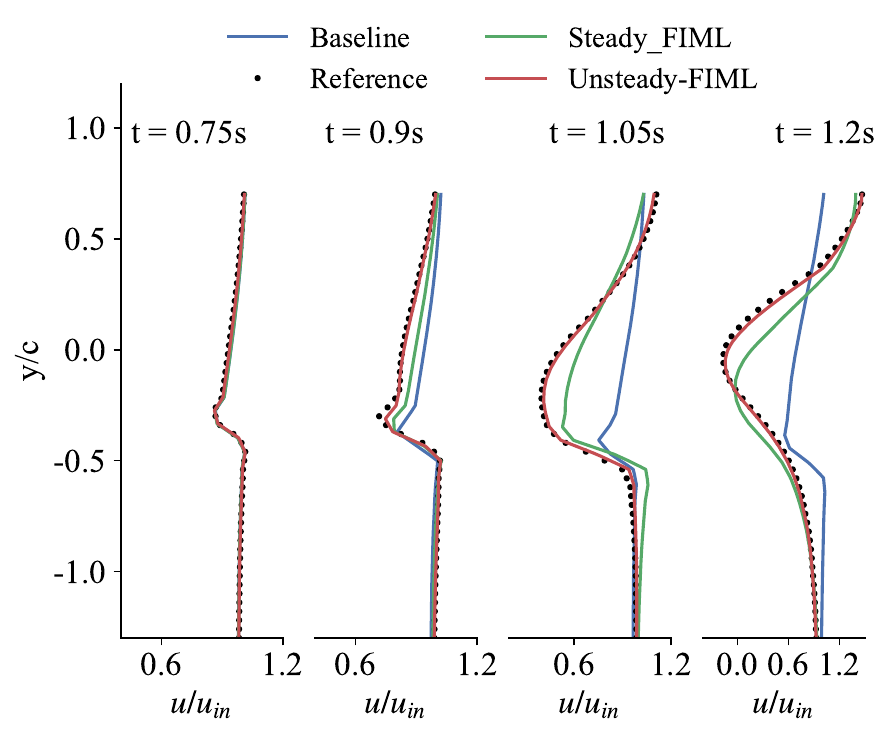}
  \caption{Velocity profile (0.5$c$ downstream) at various time instances moment among baseline, reference, and steady- and unsteady-FIML (pitch rate 0.5 rad/s).}
  \label{fig:velocity_profile-steady-unsteady}
\end{figure*}

Fig.~\ref{fig:velocity_profile-steady-unsteady} presents streamwise velocity profiles extracted at a location $0.5c$ downstream of the trailing edge at four sequential time instances during the dynamic stall. 
At $t = 0.75$ s, all models predict similar velocity profiles, indicating minimal unsteady wake development at this early stage. The profiles show a characteristic thin reverse flow region between $y/c \approx 0.25 \sim 0.45$, and show a relatively uniform flow outside this region. 
This indicates that at an early time instance, the velocity profile downstream of the trailing edge has not yet been significantly influenced by the developing separated flow over the airfoil.
By $t = 0.9$ s, the velocity profile of the baseline model begins to deviate from the reference data, which exhibits a slightly broadening and asymmetry wake profile in the region $y/c \approx 0.2 \sim 0.5$. 
However, the steady-FIML model captures the general trend of the reference profile, though it still underestimates the magnitude of the flow reverse and does not fully reproduce the reversal velocity levels observed in the reference data.
At $t = 1.05$ s, the baseline model departs substantially from the reference solution, especially in the core velocity reverse (wake) region ($y/c \approx -0.55 \sim 0.5$), where the profile exhibits significant velocity deficits.
The steady-FIML model shows reasonable improvements relative to the baseline by broadening the predicted wake and increasing the magnitude of the velocity deficit.
Nevertheless, the steady-FIML model still fails to accurately reproduce the detailed velocity profile, and the profile shape is phase-shifted relative to the reference.
By $t = 1.2$ s, the baseline model continues to substantially under-predict the wake width and velocity deficit magnitude.
Whereas the steady-FIML model demonstrates improved performance over the baseline.

These velocity profiles reveal that while the steady-FIML model provides some improvements over the baseline model during the dynamic stall event, its ability to accurately predict the velocity profiles and detailed wake structures downstream of the airfoil remains limited. 
Although the steady-FIML model consistently outperforms the baseline by producing qualitatively closer velocity profiles compared to the reference data, it still fails to achieve close quantitative agreement.
The velocity profiles and wake structures exhibit highly complex, unsteady characteristics during the dynamic stall.
This further highlights the necessity of a time-resolved unsteady-FIML model.
Again, the unsteady-FIML agrees much better with the reference than the steady-FIML, similar to what we observed in the previous figures.

\begin{figure*}[!t]
  \centering
  \includegraphics[width=0.48\linewidth]{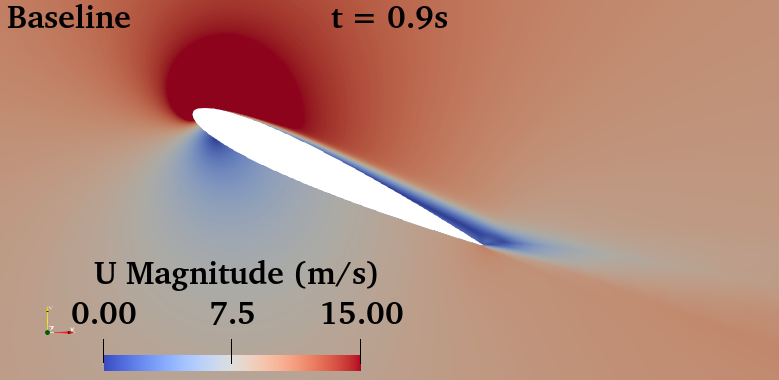}
  \includegraphics[width=0.48\linewidth]{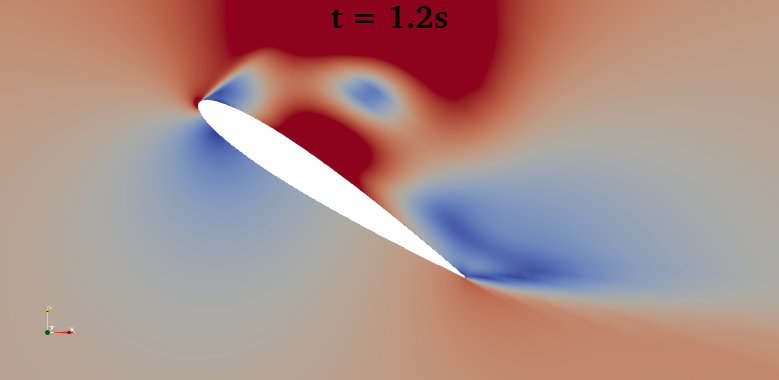}\\
  \includegraphics[width=0.48\linewidth]{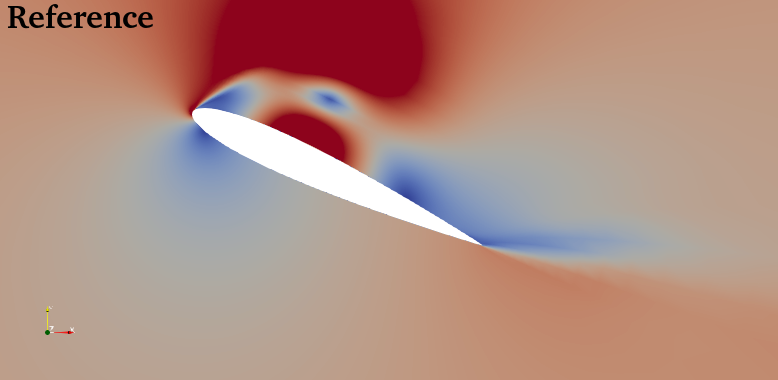}
  \includegraphics[width=0.48\linewidth]{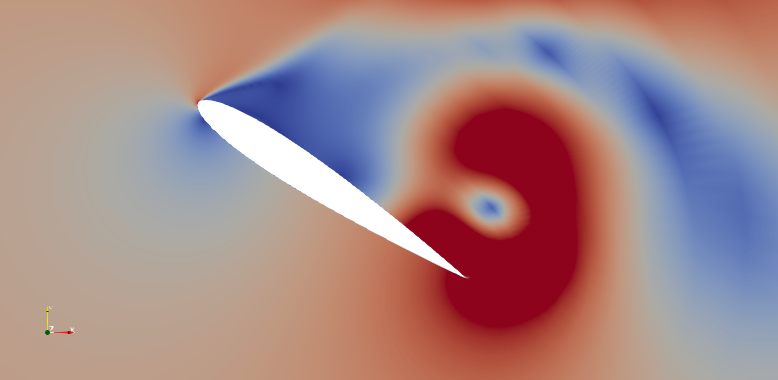}\\
  \includegraphics[width=0.48\linewidth]{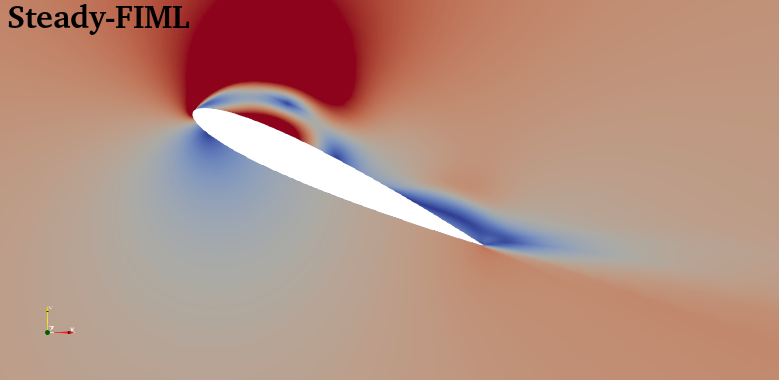}
  \includegraphics[width=0.48\linewidth]{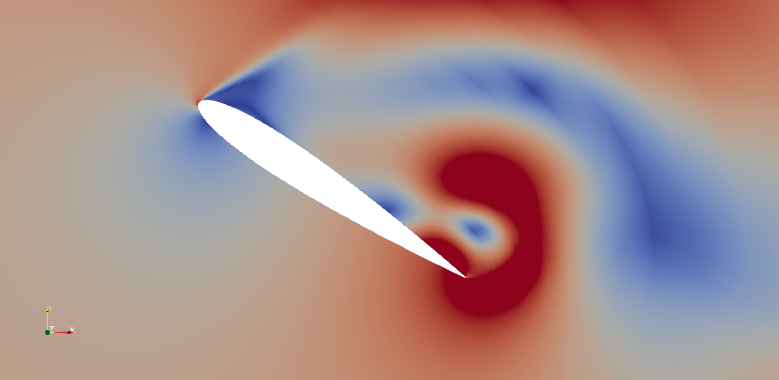}\\
  \includegraphics[width=0.48\linewidth]{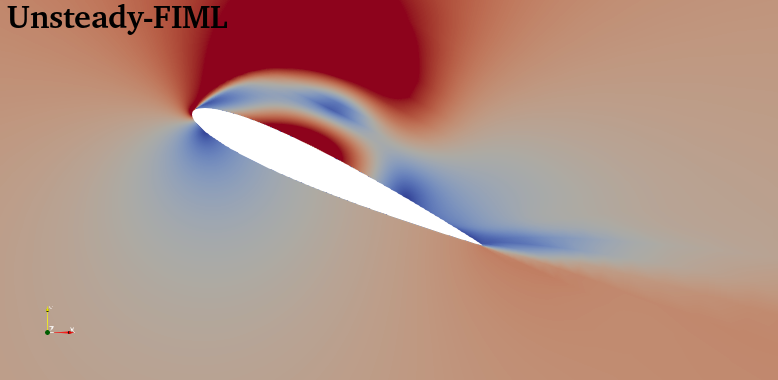}
  \includegraphics[width=0.48\linewidth]{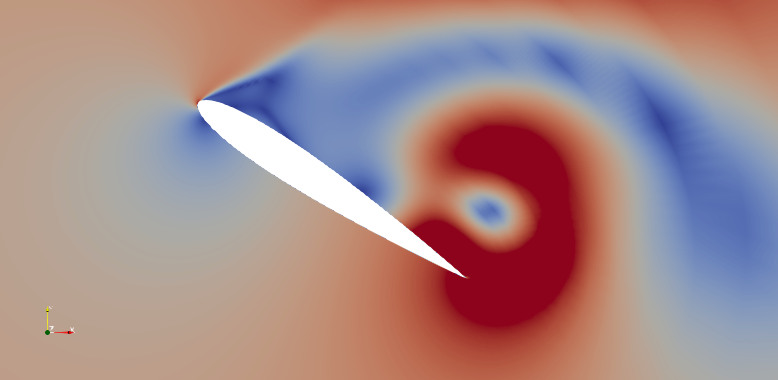}
  \caption{Comparisons of flow fields at $t = $ 0.9 (left) and 1.2 s (right) among baseline, reference, steady-FIML, and unsteady-FIML (pitch rate 0.5 rad/s).}
  \label{fig:velocity-field-predict-0.5}
\end{figure*}

Fig.~\ref{fig:velocity-field-predict-0.5} presents a comparison of instantaneous velocity fields at two different time instances and evaluates the performance of the baseline, steady-FIML, and unsteady-FIML models against the reference flow fields. 
At $t = 0.9$ s, corresponding to the early stage of dynamic stall vortex formation, the baseline model fails to reproduce the coherent vortex structures (dynamic stall vortex) near the leading edge on the upper surface of the airfoil, which is observed in the reference flow fields. 
The steady-FIML model shows moderate improvement over the baseline model, producing vortex structures near the leading edge that are smaller in size and exhibit insufficient spatial extent separation. 
The unsteady-FIML model, in contrast, shows closer agreement with the reference solution and captures the comparable size and shape of the dynamic stall vortex.
By $t = 1.2$ s, as the dynamic stall vortex develops and convects downstream, the baseline model still shows significant differences;
it fails to predict the large-scale shear layer vortex, which is clearly seen in the reference solution, at the trailing edge on the upper surface of the airfoil. 
The steady-FIML model improves upon this by capturing the vortex structure with qualitatively spatial location and general shape, but exhibits a reduced size and strength. 
On the contrary, the unsteady-FIML model demonstrates the closest agreement with the reference solution, reasonably predicting the position, size, and strength of the vortex as well as the corresponding wake pattern downstream of the airfoil. 
Overall, these comparisons again highlight the limitations of the steady-FIML model in capturing the highly transient flow phenomena. The unsteady-FIML framework, on the other hand, demonstrates an enhanced capability to reproduce the time-resolved flow physics of dynamic stall, validating its necessity for accurate unsteady aerodynamic predictions.

\subsection{Performance Evaluation of FIML Models Trained and Tested at a Different Pitch Rate (0.35 rad/s)}

In this section, we evaluate the generalizability of the steady- and unsteady-FIML models. The unsteady FIML was trained using data with a pitch rate of 0.5 rad/s. We will directly use the 0.5-pitch-rate-trained unsteady FIML model to predict dynamic stall at a pitch rate of 0.35. Since the pitch rate is irrelevant in the steady-FIML training (we use steady-state flow at different angles of attack for training), we directly use the steady-FIML trained model to predict the pitch rate of 0.35.

\begin{figure*}[!t]
  \centering
  \includegraphics[width=0.48\linewidth]{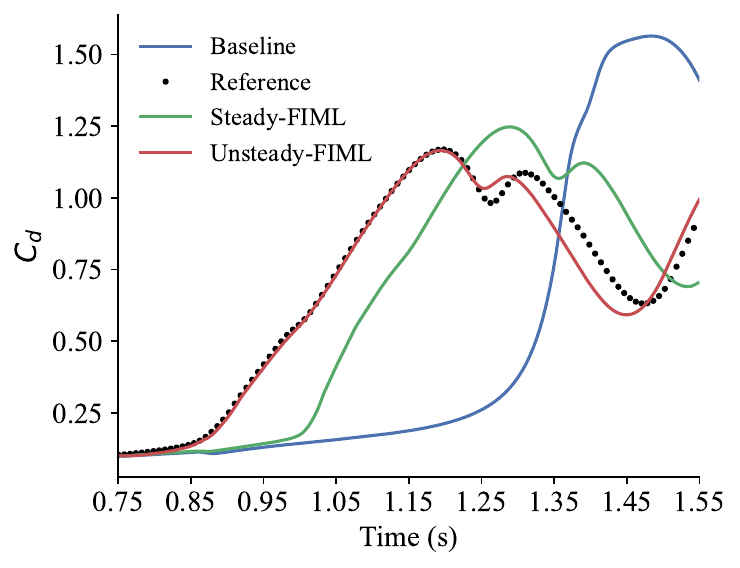}
  \includegraphics[width=0.48\linewidth]{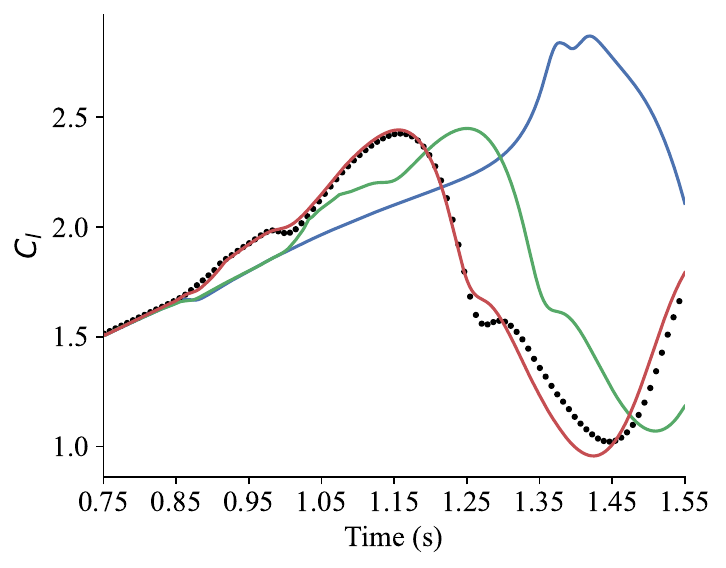} \\
  \includegraphics[width=0.48\linewidth]{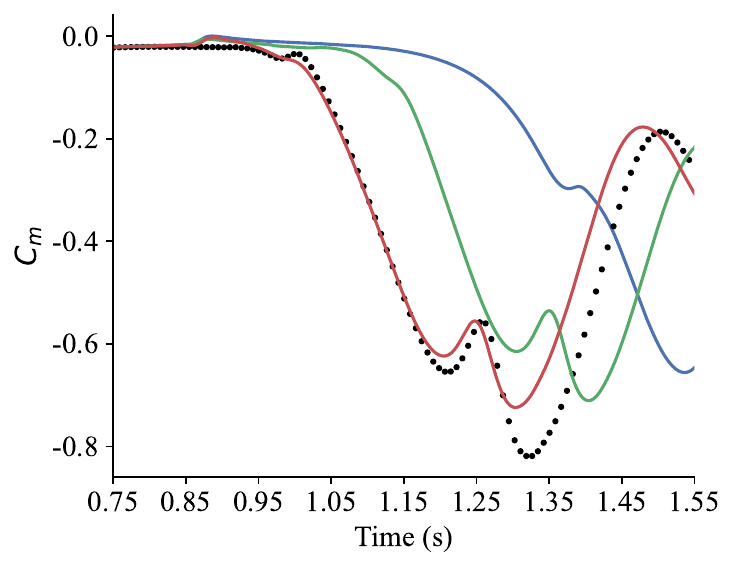}
  \caption{Temporal evolution of airfoil drag, lift, and pitching moment among baseline, reference, steady- and unsteady-FIML (pitch rate 0.35 rad/s).}
  \label{fig:using-0.5-to-predict-0.35}
\end{figure*}

Fig.~\ref{fig:using-0.5-to-predict-0.35} presents the time series of the drag, lift, and pitching moment during the dynamic stall with a pitch rate of 0.35 rad/s.
For the drag coefficient, the baseline model severely under-predicts the drag during the initial pitch-up phase, and misrepresents the peak drag magnitude and timing.
The steady-FIML model shows some improvements over the baseline and correctly initiates the drag rise earlier, but persistent discrepancies and misalignments in phase and magnitude indicate its limitations in capturing the unsteady dynamic stall physics.
In contrast, the unsteady-FIML model achieves the closest agreement with the reference data, indicating its effectiveness in capturing the complex transient behavior.
For the lift coefficient, the baseline model exhibits an over-prediction of maximum lift and a delayed stall onset relative to the reference data.
The steady-FIML model, although it corrects the maximum lift value and captures the trend of a sharp decrease in lift, exhibits a noticeable phase shift compared to the reference data.
On the contrary, the unsteady-FIML model demonstrates excellent quantitative and phase alignment with the reference data, accurately capturing the peak magnitude and the sharp lift loss.
As to the pitching moment, the baseline model demonstrates a clearly deviation from the reference data; it severely lags in capturing the pitching moment transition and underpredicts the magnitude of the negative dip.
The steady-FIML model improves upon the baseline by capturing the general downward trend.
However, it still underestimates the depth and exhibits a pronounced phase mismatch when compared to the reference data.  
Conversely, the unsteady-FIML model closely reproduces both the amplitude and phase (though slightly off the reference after $t > 1.3$ s) of the reference data. 
It accurately predicts the sharp decline in pitching moment and the characteristic double troughs.

Similar to what we found above, the steady-FIML, trained using steady-state flow data across multiple angles of attack, is unable to capture the complex, time-dependent unsteady dynamics, such as transition, separation, and vortex shedding, during the dynamic stall.
Its performance in predicting unsteady flow physics at a pitch rate of 0.35 rad/s is worse than at a pitch rate of 0.5 rad/s.
The unsteady-FIML model consistently outperforms the steady-FIML.
In addition, unsteady FIML exhibits reasonable generalizability for predicting flow conditions that lie completely outside the training parameter space, i.e., using the 0.5 rad/s pitch rate data for training and predicting a 0.35 rad/s pitch rate.

\begin{figure*}[!t]
  \centering
  \includegraphics[width=0.48\linewidth]{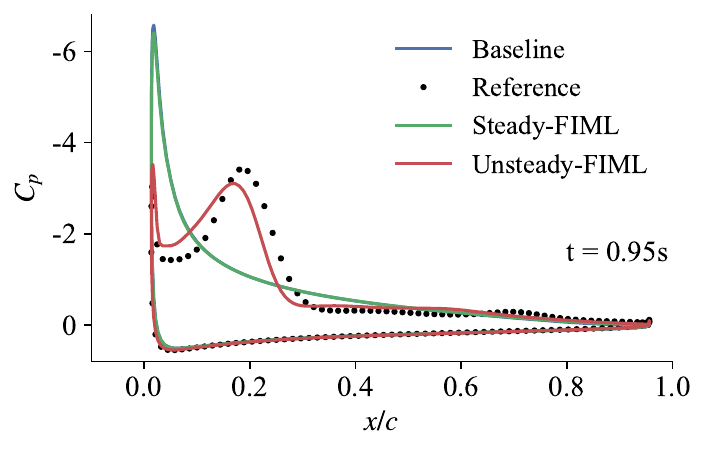}
  \includegraphics[width=0.48\linewidth]{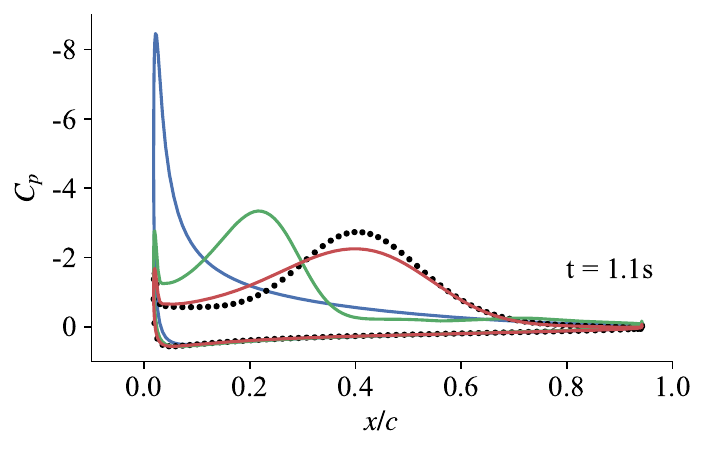} \\
  \includegraphics[width=0.48\linewidth]{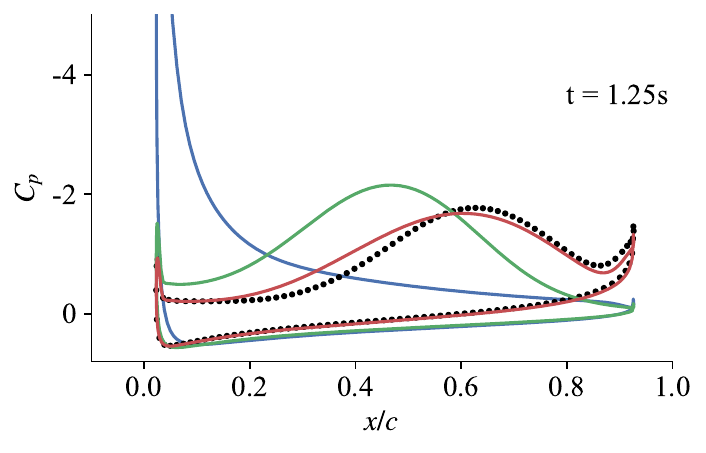}
  \includegraphics[width=0.48\linewidth]{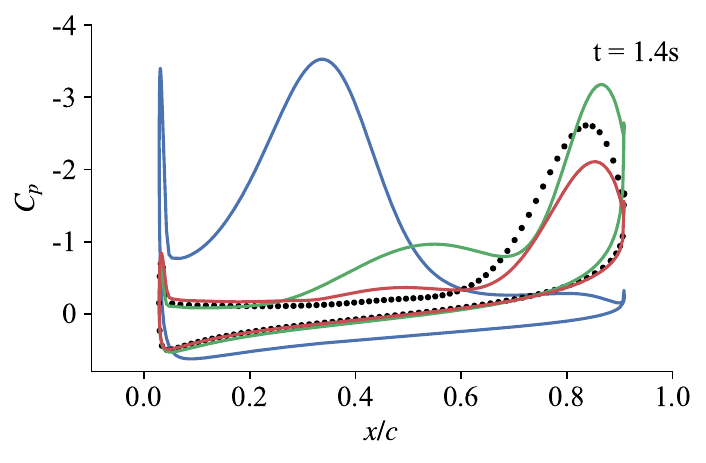}
  \caption{Surface pressure profiles at various time instances among baseline, reference, steady- and unsteady-FIML (pitch rate 0.35 rad/s).}
  \label{fig:cp-0.5-predict-0.35}
\end{figure*}

\begin{figure*}[!t]
  \centering
  \includegraphics[width=0.6\linewidth]{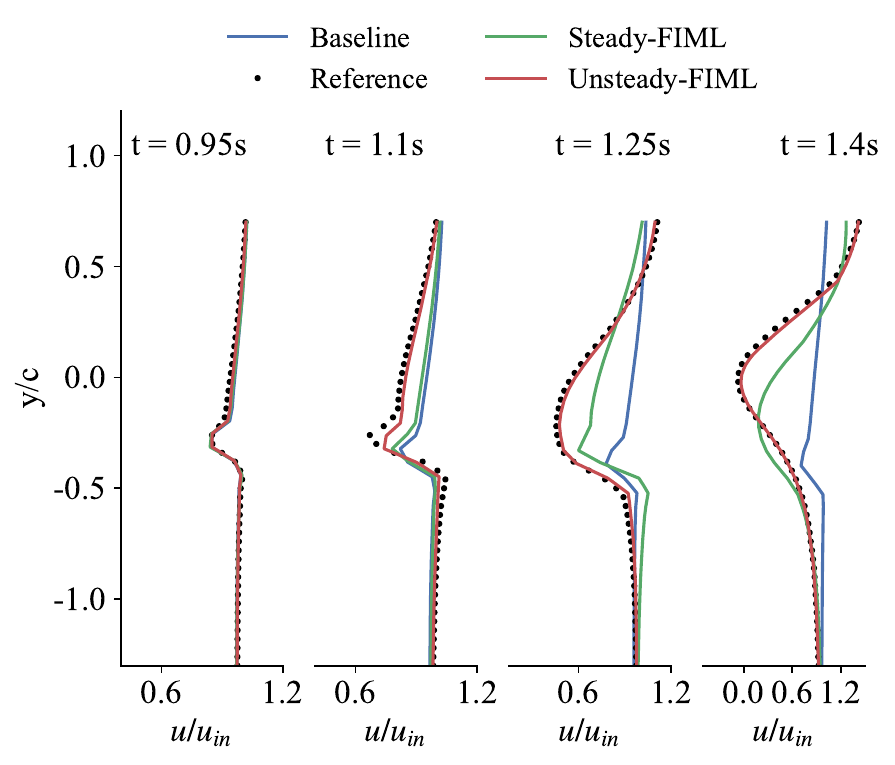}
  \caption{Velocity profile (0.5$c$ downstream) at various time instances among baseline, reference, steady- and unsteady-FIML (pitch rate 0.35 rad/s).}
  \label{fig:velocity_profile-0.5-predict-0.35}
\end{figure*}

\begin{figure*}[!t]
  \centering
  \includegraphics[width=0.48\linewidth]{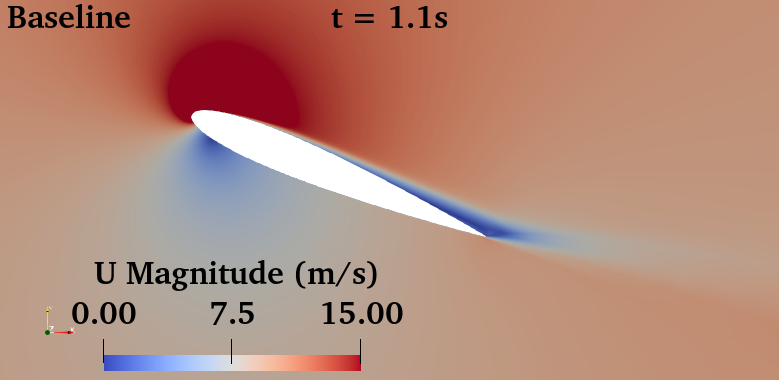}
  \includegraphics[width=0.48\linewidth]{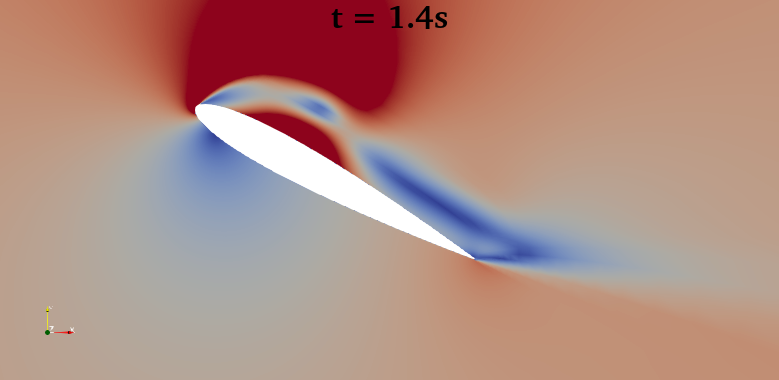}\\
  \includegraphics[width=0.48\linewidth]{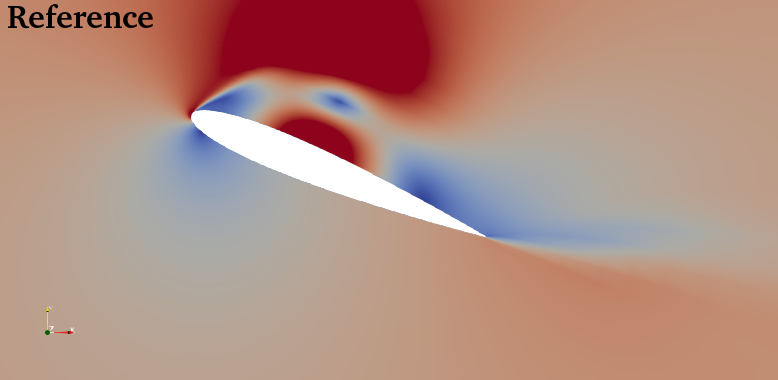}
  \includegraphics[width=0.48\linewidth]{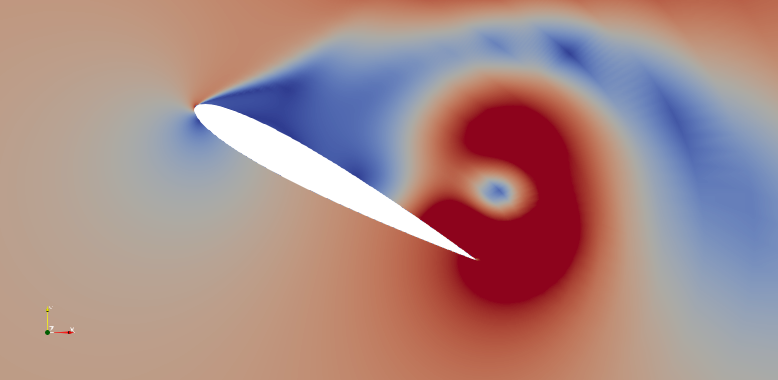}\\
  \includegraphics[width=0.48\linewidth]{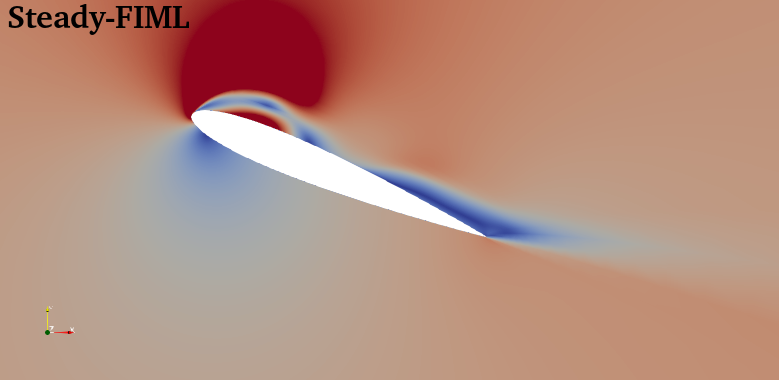}
  \includegraphics[width=0.48\linewidth]{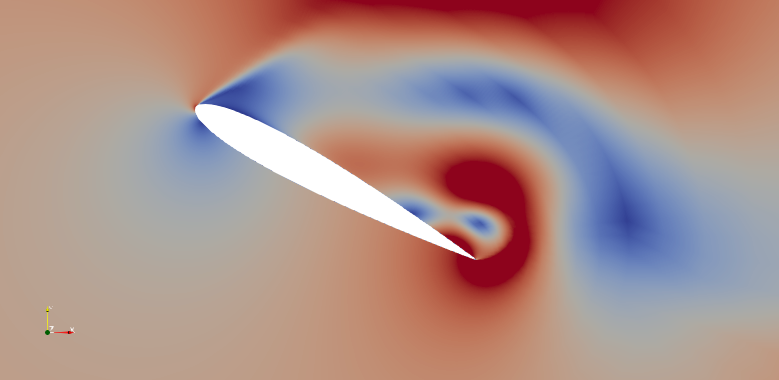}\\
  \includegraphics[width=0.48\linewidth]{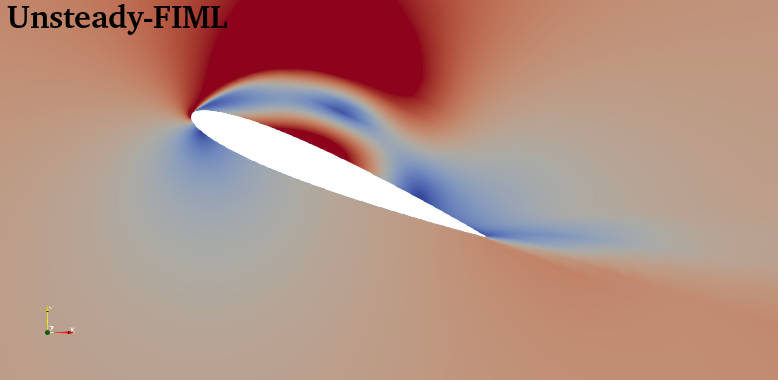}
  \includegraphics[width=0.48\linewidth]{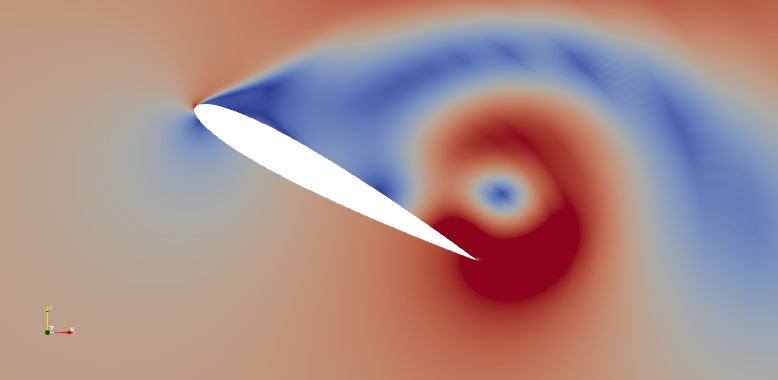}
  \caption{Comparisons of flow fields at $t = $ 1.1 (left) and 1.4 s (right) among baseline, reference, steady-FIML, and unsteady-FIML (pitch rate 0.35 rad/s).}
  \label{fig:velocity-field-predict-0.35}
\end{figure*}

We observe a similar trend in the surface pressure distribution (Fig.~\ref{fig:cp-0.5-predict-0.35}), downstream velocity profile (Fig.~\ref{fig:velocity_profile-0.5-predict-0.35}), and velocity contour (Fig.~\ref{fig:velocity-field-predict-0.35}).
Overall, the 0.5-pitch-rate-trained unsteady-FIML can successfully predict the pressure and velocity fields for the 0.35 rad/s pitch rate case, although the prediction accuracy downgrades slightly.
However, the unsteady-FIML's generalizability is much higher than the steady-state FIML.
For example, when comparing Fig.~\ref{fig:cp-steady-unsteady} with Fig.~\ref{fig:cp-0.5-predict-0.35}, and Fig.~\ref{fig:velocity_profile-steady-unsteady} with Fig.~\ref{fig:velocity_profile-0.5-predict-0.35}, the steady-FIML's errors is much larger in 0.35 rad/s pitch rate case than 0.5 rad/s.
In contrast, the unsteady FIML maintains a reasonably good accuracy, even for predicting pitching rates not used in the training.
We can also observe a similar trend in the velocity contours by comparing Fig.~\ref{fig:velocity-field-predict-0.5} right and Fig.~\ref{fig:velocity-field-predict-0.35} right.
The developed dynamic stall vortex convects downstream and induces a large-scale shear layer vortex near the trailing edge on the upper surface of the airfoil. 
The baseline model continues to exhibit delayed vortex development and fails to capture the dominant large-scale shear layer vortex. 
The steady-FIML model partially resolves the vortex structure but underestimates its shape and strength and fails to capture the correct convection behavior. 
The unsteady-FIML model, in contrast, demonstrates excellent agreement with the reference solution, reasonably predicting the vortex size, location, and wake evolution downstream.

\section{Conclusion}
\label{sec_conclusion}

This study develops an unsteady field inversion machine learning (FIML) capability to enhance RANS turbulence models for predicting time-resolved unsteady flows in airfoil dynamic stall. 
The SA turbulence model is augmented by incorporating a spatial-temporal correction field ($\beta$) into the production term, which enables the localized correction of turbulence production to better capture flow separation and time-resolved unsteady aerodynamics.
An inverse problem is solved to minimize the SA model's prediction error by optimizing the spatial-temporal distribution of the $\beta$ correction fields.
To enable generalizable prediction, a multi-layer neural network model is then trained for the relationship between the time-dependent local flow features (inputs) and the optimized $\beta$ correction fields (outputs).
Only the time series of the drag coefficient from the SST model is used as the training reference data.
For comparison, a steady-state FIML strategy is also employed, where multiple field inversions are conducted based on steady-state lift coefficients (computed from the SST model) obtained from different angles of attack.
A similar neural network model is trained that uses the optimized $\beta$ fields and local features from all the angles of attack as training data.

The time-resolved unsteady flow in a NACA0012 airfoil during pitch-up dynamic stall is used as a benchmark. 
Both steady and unsteady FIML models are evaluated by comparing their prediction accuracy for the same pitch rate used in training and a different pitch rate not used in training. 
The results show that the unsteady-FIML model achieves substantially higher agreement with the reference data compared with the steady-FIML model.
Specifically, the unsteady-FIML approach demonstrates superior performance in reproducing the spatial–temporal evolution of aerodynamic coefficients, such as drag, lift, and pitching moment, as well as surface pressure distributions and velocity fields.
It accurately captures the phase and amplitude of dynamic stall characteristics, including the onset and progression of vortex formation, the timing of stall, and the subsequent recovery, whereas the steady-FIML exhibits persistent phase shifts and magnitude discrepancies.
Instantaneous flow field comparisons reveal that the unsteady-FIML can reconstruct coherent dynamic stall vortices and wake structures that closely match those from the reference SST model, while the steady-FIML largely fails to capture the correct vortex size, position, and convection behavior.
This confirms that incorporating temporal information during training is essential for modeling the flow-history-dependent physics of dynamic stall.

Furthermore, when applied to a different pitch rate (0.35 rad/s) not used in training, the unsteady-FIML continues to demonstrate strong predictive capability and robustness, accurately reproducing unsteady aerodynamic coefficients and their temporal evolution.
This generalization test highlights the model’s ability to extrapolate beyond its training conditions; a critical requirement for practical engineering use where flow parameters vary.

Overall, the unsteady-FIML framework represents a significant advancement over traditional steady-state augmentation methods. 
By embedding the time evolution of correction fields directly into the inversion and learning process, this approach provides a physically consistent and data-efficient method to enhance RANS turbulence models for unsteady flow prediction.
The developed framework is fully integrated into the open-source DAFoam platform, providing a reusable and extensible tool for developing high-fidelity RANS turbulence models.
Future work will extend this framework by incorporating eddy-resolved simulations, such as large-eddy simulation (LES), as reference data to further improve accuracy and enable the construction of generalizable unsteady turbulence closures applicable to a wide range of flow conditions and configurations.

\section{Acknowledgments}

This material is based upon work supported by the National Science Foundation under Grant Numbers 2223676 and 2415347.
This work used the Stampede 3 supercomputer at Texas Advanced Computing Center through allocation ATM-140019 from the Advanced Cyberinfrastructure Coordination Ecosystem: Services \& Support (ACCESS) program, which is supported by U.S. National Science Foundation grants \#2138259, \#2138286, \#2138307, \#2137603, and \#2138296.
Portions of this manuscript benefited from the use of artificial intelligence tools, which were employed primarily to improve readability, grammar, and clarity of language. All intellectual content, analysis, and conclusions remain the sole work of the authors.

\bibliography{bib/refs}

\end{document}